\documentclass[12pt,preprint]{aastex}

\slugcomment{}

\shorttitle{Near-infrared photometry of Y dwarfs}
\shortauthors{Leggett et al.}

\begin{document}


\title{Near-infrared photometry of Y dwarfs: \\ low ammonia abundance and the onset of water clouds}


\author{S. K. Leggett\altaffilmark{1}}
\email{sleggett@gemini.edu}
\author{Caroline V. Morley\altaffilmark{2}}
\author{M. S. Marley\altaffilmark{3}}
\and
\author{D. Saumon\altaffilmark{4}}

\altaffiltext{1}{Gemini Observatory, Northern Operations Center, 670
  N. A'ohoku Place, Hilo, HI 96720, USA} 
\altaffiltext{2}{Department of Astronomy and Astrophysics, University of California,
Santa Cruz, CA 95064, USA}
\altaffiltext{3}{NASA Ames Research Center, Mail Stop 245-3, Moffett Field, CA 94035, USA}
\altaffiltext{4}{Los Alamos National Laboratory, PO Box 1663, MS F663, Los Alamos, NM 87545, USA}

\begin{abstract}

We present new near-infrared photometry for seven late-type T dwarfs and nine Y-type dwarfs, and lower limit magnitudes for a tenth Y dwarf, obtained at Gemini Observatory. We also present a reanalysis of $H$-band imaging data from the Keck Observatory Archive, for an eleventh Y dwarf. These data are combined with earlier MKO-system photometry, {\it Spitzer} and {\it WISE} mid-infrared photometry, and available trigonometric parallaxes, to create a sample of   late-type  brown dwarfs which includes ten T9 --- T9.5 dwarfs or dwarf systems, and sixteen Y dwarfs. We compare the data to our models which include updated H$_2$ and NH$_3$ opacity, as well as low-temperature condensate clouds. The models qualitatively reproduce  the trends seen in the observed colors, however there are discrepancies of around a factor of two in flux for  the Y0 -- Y1 dwarfs, with 
$T_{\rm eff} \approx$ 350 -- 400~K. At $T_{\rm eff} \sim 400$~K, the problems 
could be addressed by significantly reducing the NH$_3$ absorption, for example by halving the abundance of NH$_3$ possibly by vertical mixing.
At $T_{\rm eff} \sim 350$~K, the discrepancy may be resolved by incorporating thick water clouds. The onset of these clouds might occur over a narrow range in $T_{\rm eff}$, as indicated by the observed small change in 5 $\mu$m flux over a large change in $J - W2$ color. Of the known Y dwarfs, 
the reddest in $J -$W2 are  WISEP J182831.08$+$265037.8 and WISE J085510.83$-$071442.5. We interpret the former as a pair of identical 300 -- 350~K dwarfs, and the latter as a 250~K dwarf. If these objects are $\sim 3$ Gyrs old, their masses are $\sim$10 and $\sim$5 Jupiter-masses respectively. 

\end{abstract}


\section{Introduction}

In 1995 the first definitive detection of an exoplanet orbiting a main-sequence star was announced (Mayor \& Queloz 1995). In the same year, and in fact in the same edition of {\it Nature},  the first  definitive detection of a brown dwarf (an object with  insufficient mass for stable hydrogen-burning) was also announced (Nakajima et al. 1995). The detection of exoplanets continued at a rapid rate. However it was not until 1999 that more  brown dwarfs were found, in the early data releases of the Sloan and 2MASS sky surveys (Burgasser et al. 1999, Strauss et al. 1999). The first exoplanet and brown dwarf discoveries were of relatively warm and massive sources. The exoplanets were of the ``hot Jupiter'' class, Jupiter-mass objects close to their star. The brown dwarfs were what are now known as mid-T class objects, with effective temperatures ($T_{\rm eff}$) $\approx 1000$~K, and mass $\approx 50$~Jupiter-masses. Exoplanets that are much less massive and further from their host stars are now known, for example the planetary system around Kepler 11, with six planets at 0.1 -- 0.5 AU, and masses as low as 2 Earth-masses (Lissauer  et al. 2011). Similarly, cooler and lower mass brown dwarfs are now known (Cushing et al. 2011). The two populations have a significant degree of overlap, and brown dwarfs can show many observational similarities to directly-imaged exoplanets (Liu et al. 2013). There is active debate on how the formation mechanisms differ  and relate to each other (e.g. Beichman et al. 2014 and references therein).

This paper continues our series of papers where we present new observations of brown dwarfs, and compare   these data to state-of-the-art models calculated by members of the group. The known brown dwarf population has been extended to intrinsically fainter sources by sky surveys, most recently the Wide-field Infrared Survey Explorer ({\it WISE}; Wright et al. 2010). The models
have become increasingly advanced using new  pressure- (or collision-)induced H$_2$ absorption and NH$_3$ opacity (Saumon et al. 2012), and incorporating low-temperature condensate cloud decks (Morley et al., 2012, 2014). Our earlier papers include the verification of the reddening caused by silicate, sulphide and chloride clouds in L-, T- and Y-type dwarf atmospheres (Stephens et al. 2009, Leggett et al. 2013). Here we present new near-infrared photometry, and compare our dataset to models which include water clouds (Morley et al. 2014).

\section{Observations}

In this paper we present new near-infrared photometry for seven late-type T dwarfs and nine Y-type dwarfs,
obtained using the Gemini Observatory's Near-Infrared Imager (NIRI, Hodapp et al. 2003) and FLAMINGOS-2 (Eikenberry et al. 2004). For a tenth Y dwarf, we present a reanalysis of $H$-band imaging data taken with the Keck Observatory's NIRC2 near-infrared imager. We start this section with the presentation of new upper-limit fluxes  (lower-limit magnitudes) for an eleventh Y dwarf.

\subsection{$zYJ$ limits for WD  0806$-$66B using GMOS-South and GSAOI}

We observed WD  0806$-$66B,  a very late-type brown dwarf companion to a white dwarf, at Gemini South using
the Gemini Multi-Object Spectrograph (GMOS, Hook et al. 2004) and the Gemini South Adaptive Optics Imager (GSAOI, McGregor et al. 2004 and Carrasco et al. 2012). The primary,   WD  0806$-$66, is also referenced in the literature as L 97-3, LTT 3059, NLTT 19008 and GJ 3483. While to our knowledge the label ``WD'' has never been used for an object that is not a white dwarf, this brown dwarf companion has been mainly referred to in the literature as  WD  0806$-$66B, and we adopt that label here. 

The brown dwarf
was discovered in a search for common proper motion companions using {\it Spitzer}
4.5~$\mu$m images (Luhman, Burgasser \& Bochanski 2011). Follow-up  {\it Spitzer} 3.6~$\mu$m imaging and deep $J$-band imaging confirmed the source to be one of the coldest brown dwarfs known, with $T_{\rm eff} \approx 350$~K (Luhman et al. 2012). Given the constraints on age imposed by the white dwarf, the mass of WD  0806$-$66B is $< 13$ M$_{\rm Jup}$ (Rodriguez et al. 2011). 

Deeper 1~$\mu$m imaging using the F110W filter and the Wide Field Camera 3 (WFC3) on the {\it Hubble Space Telescope (HST)} has now provided the first detection of the source in the near-infrared (Luhman et al. 2014b). Luhman et al. find 
$m_{110W} = 25.70 \pm 0.05$, and use observed spectra for T9 to Y0.5 dwarfs, as well as model-generated spectra, to estimate $m_{110W} - J \sim 0.7$. We have confirmed and further constrained $m_{110W} - J$ 
 using observed spectroscopy of the same three brown dwarfs used by Luhman et al. 
(UGPS J072227.51$-$054031.2 [UGPS 0722, T9], WISE J014807.25$-$720258.7 [W0148, T9.5], WISE J154151.65$-$225024.9 [W1541, Y0.5]),
and adding data for the resolved T9 and Y0 system WISEPC J121756.91+162640.2A and B (W1217A,B; Leggett et al. 2014). We find that for the three T9 -- T9.5 dwarfs $m_{110W} - J = 0.84 \pm 0.06$ and for the two Y0 -- Y0.5 dwarfs  $m_{110W} - J = 0.71 \pm 0.09$. \footnote{We investigated trends with spectral type for $m_{110W} - Y$ but found more scatter, perhaps due to the fact that spectral type is based on the shape of the $J$-band flux peak,  and  $Y - J$ is sensitive to gravity, metallicity and cloud properties (\S 7).  Burgasser et al. (2006) give NICMOS F110W magnitudes for a sample of T dwarfs, and those data imply $m_{110W} - J = 1.1 \pm 0.1$. However the instrument handbooks show that the system throughput of the NICMOS $+$ F110W filter is significantly different from the WFC3  $+$ F110W filter.} We exclude model spectra, given the discrepancies between  observations and theory (\S 6). The mid-infrared properties of  WD  0806$-$66B are similar to the Y0.5 W1541 (\S 4), hence we adopt $J = 25.00 \pm 0.10$ for the source.

The GMOS observations were executed via program GS-2011B-Q-53 on UT 2011 December 5, 30 and 31, and 2012 January 19, 20 and 21.
Forty-five 720-second exposures at $z$ were obtained for a total on-source time of 9 hours. Conditions were photometric  with seeing around $0\farcs 8$. Landolt (1992) and Smith et al. (2002) photometric standards were used for calibration each night, as is routine practice at Gemini. The data were reduced in the usual way using IRAF routines, paying particular attention to fringe correction. 

The GSAOI observations were executed via program GS-2013B-Q-15. $Z$-band observations were obtained on UT 2013 December 17 and 18, and $J$-band on UT 2013 December 18 and 2014 January 15, 16 and 19.  The total on-source time was 94 minutes at $Z$ and 91 minutes at $J$, consisting of 120-second exposures at $Z$ and 60-second at $J$. Photometric but poorer natural seeing conditions  prevailed, and  the adaptive optics corrected images had full-width half-maximum (FWHM) of  $0\farcs3$ -- $0\farcs4$. Stacking the images was difficult as there are relatively few point sources, and the FWHM of the stacked image was degraded further to 
$\sim 0\farcs5$. The T9 brown dwarf    UGPS 0722
and other sources near this T9 (in the $85\arcsec$ by $85\arcsec$ GSAOI field of view), were imaged on 2014 January 19, and used to determine zeropoints at $Z$ and $J$ and to investigate whether the GSAOI photometric system is equivalent to the Mauna Kea photometric system (MKO, Tokunaga \& Vacca 2005). We also investigated the linearity behavior. We found that no linearity correction needed to be applied, and that the GSAOI $Z$ and $J$ filters correspond directly to the MKO $Y$ and $J$.   We derived identical zeropoints, within the uncertainties, for both bright and faint read modes and for each of the four detectors. The zeropoints were measured to be  26.71 at  GSAOI-$Z$ (equivalent to MKO-$Y$) and 26.40 at $J$, where zeropoint is defined  as the magnitude of an object that produces one count (or data number) per second. 

Figure 1 shows our Gemini images with the location of the source indicated. 
We used the 4.5~$\mu$m image to derive coordinates for  WD  0806$-$66B of 8h 07m 15.195s $-66\degr 18\arcmin  51\farcs 25$ at epoch 2009 August 24. The proper motion of the system (as determined from the white dwarf)  is $0\farcs 3403$ yr$^{-1}$ in Right Ascension and $-0\farcs 2896$ yr$^{-1}$ in Declination, and the trigonometric parallax is $52.17 \pm 1.67$ mas   (Subasavage et al. 2009).  
WD  0806$-$66B is not detected in our data, implying $z_{AB} > 26.2$, $Y > 23.5$ and  $J > 23.9$, where these values correspond to the $3 \sigma$ limits in the stacked $z$, $Y$ and $J$ images.  

Our $z$ and $Y$ data extend the limits found by Luhman et al. (2014b), using archived VLT data, to fainter magnitudes. Note that our models imply that $z - J$ turns to the blue for $M_J > 18$, or T8 and later types. 
For early-type Y dwarfs  $z - J \sim 3$ as opposed to  $z - J \sim 4$ for late-type T dwarfs. Detection in the $z$-band of such cool objects however remains challenging, as $J$ is intrinsically faint.  
Our $J$ limit is consistent with the $J = 25.0 \pm 0.1$ implied by the  $m_{110W} = 25.7 \pm 0.05$ measured by Luhman et al.  
Using the archived  {\it Spitzer} data we determine [3.6] $= 19.28 \pm 0.10$ and [4.5] $= 16.78 \pm 0.05$ (slightly different from the values derived by Luhman et al. 2012). The  WISE  All--sky source catalog gives W2 (4.6 $\mu$m) $= 17.68 \pm 0.41$, W3 (12 $\mu$m) $= 12.54 \pm  0.16$ and W4  (22 $\mu$m)$= 10.18 \pm  0.53$, however the W3 and W4 fluxes are much too bright, likely due to the bright infrared cirrus coincident with the source that can be seen in the WISE images. 

The near-infrared magnitude limits determined here are given in Table 1. 
No near-infrared spectrum is available for this brown dwarf, and so a spectral type cannot be derived.
For plotting purposes 
we assign a nominal spectral type of Y1, based on its similarity to Y0.5 -- Y1 dwarfs in the color-color diagrams shown later.
We compare the near- to mid-infrared colors of WD 0806$-$66B  to model calculations in  \S 6.

\subsection{NIRI Photometry}

NIRI was used on Gemini North to obtain $YJHK$ photometry, or a subset  thereof, for eight T8 -- Y0.5 dwarfs via programs GN-2013A-Q-63 and GN-2013B-Q-27.
The photometry is on the MKO system, however there is some variation in the $Y$ filter bandpass between the cameras used on Mauna Kea, and  $Y_{\rm NIRI} - Y_{\rm MKO} = 0.17 \pm 0.03$ (Liu et al. 2012).
Exposure times of 30~s or 60~s were used, with a 5- or 9-position telescope dither pattern. The data were reduced in the standard way using routines supplied in the IRAF Gemini package. UKIRT Faint Standards  were used for calibration, taking the $Y$ data from the UKIRT online catalog\footnote{http://www.jach.hawaii.edu/UKIRT/astronomy/calib/phot cal/fs ZY MKO wfcam.dat} and the $JHK$ data from Leggett et al. (2006). All data were taken on photometric nights with typical near-infrared seeing of  $0\farcs 8$. 

Table 1 gives the results, where we have applied the correction above to the $Y$ data to get it on the standard MKO system. The date of observation is also listed. Note that the $H$-band measurement for W1541 is a repeat of the measurement we published in Leggett et al. (2013), and is significantly brighter ---
here we find $H = 21.07 \pm 0.07$ compared to the previous value of $H = 22.17 \pm 0.25$.
The data used here were obtained in much better seeing, $0\farcs5$ compared to $0\farcs7$, which allowed better separation of the target from a very close star. Hence we use the results from the more recent dataset only.

\subsection{FLAMINGOS-2 Photometry}

FLAMINGOS-2  was used on Gemini South  to obtain $YJHKs$ photometry, or a subset thereof, for eight T9 -- Y1 dwarfs, via programs GS-2013B-Q-15, GS-2014A-Q-50 and GS-2014B-Q-17. In order to confirm that the photometric system corresponds to MKO for the $YJH$ filters, and to determine the transformation from $Ks$ to $K$, we observed 
2MASS J04151954$-$0935066 (T8)  and  UGPS 0722 (T9)
on 2013 November 22 and 27, and December 25.
For the science targets, exposure times of 60~s or 120~s were used for $Y$, 60~s for $J$, and 10~s or 15~s for $H$ and $Ks$, with a 5- or 9-position telescope dither pattern. The data were reduced in the standard way using routines supplied in the IRAF Gemini package. The instrument was inadvertently used in both bright and faint read mode, so zeropoints were derived for both modes. We derived zeropoints at $Y/J/H/Ks$ of 
24.83/24.91/25.15/24.45 and  24.09/24.16/24.39/23.68 for bright and medium mode respectively.

The FLAMINGOS-2 $Ks$ filter profile is very similar to that of the 2MASS $Ks$ filter. Stephens \& Leggett (2004) find that $K - Ks  = 0.15$ for T8 spectral types. We determine a slightly larger correction for FLAMINGOS-2, measuring  $K - Ks  = 0.3$ for  2MASS 0415 and   $K - Ks  = 0.5 \pm 0.1$ for UGPS 0722. For all targets presented here we adopt  $K - Ks  = 0.4 \pm 0.1$. The $YJH$ photometry was found to be on the MKO system, as expected from the filter bandpasses.

Table 1 gives the results, where we have applied the correction above to the $Ks$ data to transform it to MKO $K$. The date of observation is also listed. The upper flux limits quoted at $H$ and $K$ for WISE J064723.23$-$623235.5 (W0647) correspond to the $3 \sigma$ detection limits in each dataset.

\subsection{Keck Observatory NIRC2 Archived Data}

We used the  Keck Observatory Archive (KOA) to access $H$-band imaging data for  WISE J182831.08$+$265037.7 (W1828), taken with the NIRC2 near-infrared imager. 
The spectral sub-type of W1828 is uncertain as the near-infrared spectrum of this faint object is noisy. In the literature it has been labelled as $\geq Y2$ (e.g. Beichman et al. 2014). Here we classify it as Y1.5 as it is not very different from the Y1 dwarfs in color-magnitude diagrams. Note that W1828 is more luminous than, and is therefore assumed to be 
earlier in type than, WISE J085510.83$-$071442.5 (Luhman 2014, W0855). There is no spectrum of W0855, and  it has been assigned a type $>Y2$ in the literature based on the type assigned to W1828 (e.g. Tinney et al. 2014). Here we assign a nominal spectral type of Y2 to W0855 for plotting purposes (but see discussion in \S 4).
We used NIRC2 data taken on 2011 October 16  for  principal investigator Beichman. The night was photometric with excellent $0\farcs3$ seeing. Thirty-three frames were obtained, each consisting of a 60-second times 2-coadd exposure. A NIRC2 $H$-band flat was also downloaded from the KOA site. The data were reduced in the standard way using IRAF routines, and stacked using offsets measured from the images. The stacked image was calibrated using 10 sources in common with a shallower image of the field obtained using NIRI  on 2014 July 02, via program GN-2014A-Q-64. The NIRI image in turn was calibrated using UKIRT photometric standards; both NIRC2 and NIRI use MKO $H$ filters. 

We derive $H = 22.73 \pm 0.13$ for W1828, somewhat fainter ($1.6 \sigma$) than the 22.42$\pm$ 0.14 determined by Beichman et al. (2013) from the same dataset using different secondary standards, and consistent with the  22.85 $\pm$ 0.24 measured by Kirkpatrick et al. (2012) using 2010 NIRC2 data. In this paper we adopt our measurement of $H = 22.73 \pm 0.13$.

\section{Photometry Compilation and Discrepancies Between Published Magnitudes}

Discrepancies between near-infrared photometry obtained at Gemini and obtained by the WISE team have been noted previously (e.g. Leggett et al. 2013). Figure 2 compares $J$ and $H$ magnitudes obtained by the two groups. Gemini data are taken from this work and Leggett et al. (2013); WISE team data are from Beichman et al. (2013, 2014), Kirkpatrick et al. (2012, 2013) and Mace et al. (2013).  The sample consists of 14 T8 to Y1 dwarfs where we have data in common. It can be seen that for about one-third of the sample the two measurement sets differ by more than $2 \sigma$, with two $J$ measurements differing by 6 to 7 $\sigma$.
Variability in near-infrared flux has been observed for brown dwarfs. Primarily this is seen for late-L and early-T dwarfs, and has been associated with the transition from cloudy to clear atmospheres (e.g. Radigan et al. 2014). Clouds are also present in the atmospheres of late-T and Y dwarfs, however
the level of variability observed for the warmer brown dwarfs is typically  $<< 10$\%, and it seems unlikely that variability can explain the large discrepancies of up to   one magnitude seen here. As the scatter is larger for $J$ than $H$, and there are generally more variations in $J$ filter bandpasses used at observatories  than in $H$, we suspect the cause is unrecognised differences in photometric systems. These differences can be very large for objects with unusual energy distributions, such as T and Y dwarfs (e.g. 0.4 magnitudes for mid-T types, Stephens \& Leggett 2004).
The trends with spectral type shown in \S 4 support the Gemini measurements and not the WISE team results, where the two datasets disagree. Examples include the T9 0005$+$43 for which Mace et al. give $J - H = 0.23 \pm 0.14$ compared to our $-0.39 \pm 0.03$, and the T9pec 0146$+$42 (Dupuy et al. 2015) for which Kirkpatrick et al. (2012) and Beichman et al.  give $J - H = -1.51 \pm 0.33$ compared to our   $-0.61 \pm 0.14$. 

To avoid possible systematic errors, we use only near-infrared photometry for the Y dwarfs that we are confident is on the MKO system, in this analysis. The exceptions are the recent $J$-band detections of  WD 0806$-$66B (Luhman et al. 2014b) and W0855 (Faherty et al. 2014). We include these objects in order to populate the low-luminosity end of the sample. The WFC3 detection of WD 0806$-$66B is described in \S 2.1. For W0855,   Faherty et al. used the FourStar imager at Las Campanas Observatory and measured $J3 = 24.8^{+0.53}_{-0.35}$,  implying  $J_{\rm MKO} = 25.0^{+0.53}_{-0.35}$. We adopt the parallax given by Luhman \& Esplin (2014) and assign a nominal spectral type of Y2 to this source for plotting purposes (see \S 2.4). 
Table 2 gives $YJHK$, or a subset thereof, for 17 Y dwarfs with MKO-system near-infrared photometry, together with spectral types and distance moduli as derived from published parallaxes. 

For this paper, T and Y dwarf 
photometry  is from this work, the UKIRT Infrared Deep Sky Survey (Lawrence et al. 2007), the AllWISE catalog (Wright et al. 2010, Mainzer et al. 2011), Burningham et al. (2010, 2013), Chiu et al. (2006), Dupuy et al. (2014), Faherty et al. (2014), Knapp et al. (2004), Leggett et al. (2007, 2010, 2013), Liu et al. (2011, 2012), Lucas et al. (2010), Luhman et al. (2014b), Patten et al. (2006), Pinfield et al. (2014a,b), and Wright et al. (2014). 
Trigonometric parallaxes for the sample are taken from the Hipparcos catalog (ESA 1997), Beichman et al. (2014), Dupuy \& Kraus (2013), Luhman \& Esplin (2014), 
Manjavacas et al. (2013), Marsh et al. (2013), Subasavage et al. (2009), Tinney et al. (2014), 
Vrba et al. (2004), and Wright et al. (2014).

\section{Observed Trends with Spectral Type}

Figure 3 shows absolute 1 -- 5 $\mu$m magnitudes as a function of spectral type, and Figure 4 shows 0.9 -- 12.0 $\mu$m colors as a function of type, for a sample of late-T and Y dwarfs. New data presented here are indicated by red symbols. Color sequences are better defined, as is the intrinsic scatter in the sequences, when the new data are included. The new data enable a more meaningful comparison with the models (\S 6).

Currently the Y dwarf spectral classification is defined by the width of the $J$-band flux peak in the (often noisy) near-infrared spectrum (Cushing et al. 2011, Mace et al. 2013). There are no pronounced near-infrared spectral differences between T and Y dwarfs, for the current sample of known objects. It is quite likely that as more Y dwarfs are discovered, and our understanding of their physics and chemistry improves, the late-T and Y dwarf spectral classification scheme will have to be revised (even though spectral classification should be driven by spectral features and not by physical interpretation of the atmospheres).

In the current classification scheme, $T_{\rm eff}$ across the T/Y boundary drops sharply --- from 750~K at T8,
to 600~K at T8.5,  to 500~K at T9, to  400~K at Y0 (e.g. Saumon et al. 2007, Smart et al. 2010, Leggett et al. 2012 and 2013). The atmospheres are also changing dramatically, from being reasonably clear for mid-T types, to cloudy for late-T and early-Y with sulphide and chloride condensates, to cloudy with water ice for later Y types (Morley et al. 2012, 2014).  

Outliers are identified in Figures 3 and 4. In the case of WISE J035000.32$-$565830.2 (W0350) and WISEP J140518.40$+$553421.4 (W1405), the brightness and color trends  would be improved if their classifications were shifted 0.5 subclass later:  W0350 to Y1.5 and W1405 to Y0.5. 
However for W0350 the absolute magnitude-color relations shown later suggest that the absolute magnitudes are also too faint, in other words the distance is too small and measured parallax too large. For one source, WISE J053516.80$-$750024.9 (W0535), both the near- and mid-infrared absolute magnitudes appear too bright, implying that the distance is too large (parallax too small)  or the source is multiple.  

W0855 is extremely red in $J -$ W2, and is intrinsically  fainter than any  known Y dwarf, by  $\sim$4 magnitudes at $J$ and $\sim$2 magnitudes at W2 (neglecting W0350 whose parallax we suspect is erroneous). More Y dwarfs with good quality near-infrared spectra are required to establish the classification scheme beyond Y1 types.
A simple extrapolation of the $M_{\rm W2}$ as a function of type diagram suggests a spectral type for W0855 of around Y4.

Other sources appear to have unusual spectral energy distributions.  SDSS J141624.08$+$134826.7 (S1416B)
WISE J033515.01$+$431045.1 (W0335)  and WISE J014656.66$+$423410.0 (W0146) appear to be 
unusually faint in the $K$-band and blue in $J - K$ and $H - K$. S1416B is a known low-metallicity high-gravity T dwarf (e.g. Burningham et al. 2010).
Similarly, Beichman et al. (2014) show that W0335 has quite a  large tangential velocity, and their model interpretation suggests a relatively high gravity and an age around 8 Gyr. 
WISE J182831.08$+$265037.7 (W1828) appears unusually bright in the mid-infrared, and has been widely discussed in the literature (e.g. Beichman et al. 2013); we return to this object later in \S 7.4.

We also see an indication of  populations with unusual $Y - J$ colors. ULAS J232600.40$+$020139.2 (U2326), WISE J041358.14$-$475039.3 (W0413) and WISE J073444.02$-$715744.0 (W0734) appear to be red in $Y - J$, and 
WISEP J173835.53$+$273258.9 (W1738) and 
W0350 appear blue in this color. We discuss $Y - J$ colors in \S 7.3.

\section{Description of the Models}

In this analysis we use spectra and colors generated from model atmospheres with a variety of cloud cover. We include cloud-free models from Saumon et al. (2012) and models with homogeneous layers of  chloride and sulphide  clouds from Morley et al. (2012). We also use the  patchy cloud models from Morley et al. (2014), which include water clouds and have a surface that is partly clear and partly cloudy. These patchy cloud models  assume a surface cloud cover fraction of 50\%, and   a relatively high sedimentation efficiency described by $f_{\rm sed} = 5$, i.e. thin cloud decks. For this paper we have generated models with water clouds that have a larger surface coverage fraction of 80\%, and thicker cloud decks parameterised by  $f_{\rm sed} = 3$. All of the models have solar metallicity and are for the equilibrium chemistry case, i.e. do not include vertical mixing in the atmosphere. 
The models have been generated for surface gravities of $\log g =$ 4.00, 4.48 and 5.00 cm/s$^{-2}$.  
For a (necessarily) nearby sample of brown dwarfs with $300 < T_{\rm eff} < 600$~K and expected local-neighborhood ages between 0.4 and 10 Gyr, evolutionary models show that $4 \lesssim \log g \lesssim 5$ (Saumon \& Marley 2008).  \footnote {The magnitudes generated from the published models and used in this paper can be obtained at http://www.ucolick.org/~cmorley/cmorley/Models.html}

Morley et al. (2014) show that for temperatures between 300~K and 800~K, the expected dominant  molecular  opacity species are: CH$_4$, H$_2$, H$_2$O, NH$_3$ and PH$_3$. 
The models include updated opacities for NH$_3$ and pressure-induced H$_2$  (Saumon et al. 2012). The CH$_4$ line list has not been updated and is known to be incomplete, however at the low temperatures of this sample the incompleteness is not expected to be severe and the effect could be modest.
Figure 5 shows the opacity cross sections as a function of wavelength,  extracted from the  50/50 $f_{\rm sed} = 5$ patchy water cloud model  with $T_{\rm eff} = 400$~K, $\log g =$ 4.48 (typical of a Y0 dwarf, Leggett et al. 2014). The cross sections have been calculated for the atmosphere at the pressure and temperature of three different layers: that from which the 1 $\mu$m flux emerges, 
that from which the 5 $\mu$m flux emerges, and that from which the 10 $\mu$m flux emerges. The near- to mid-infrared flux emerges from layers with very different pressures and temperatures, and Figure 5 shows the dominant opacity sources for each of the three bandpasses.

Figure 6 shows  synthetic 0.8 -- 20  $\mu$m  spectra for a  $\log g =$ 4.48 brown dwarf at 10~pc, with different cloud cover parameters and $T_{\rm eff} = 400$~K, 300~K and 250~K. These temperatures and gravity correspond to an approximately 10 Jupiter-mass object aged 3 to 10~Gyr.  
Far-red, near- and mid-infrared filter bandpasses are also shown in Figure 6.   The water ice condensates scatter at wavelengths $\sim 1 \mu$m, and absorb at  $\sim 5 \mu$m (Morley et al. 2014, their Figure 2). Figure 6 shows that when the water cloud extent and depth is increased, the $z$ and W2 fluxes are decreased, and the flux is redistributed to $K$, [3.6] and W3. At   $T_{\rm eff} = 400$~K the flux at $Y$ and $J$ also decreases with increasing water clouds. At  $T_{\rm eff} = 300$~K and 250~K very little flux emerges in the near-infrared, independent of the cloud parameters. 

We compare the cloud-free and cloudy models to the observations in the next section.

\section{Comparing Data to Models}

Figures 7 and 8 are near- and mid-infrared color-magnitude diagrams. Figure 9 shows
near-infrared colors against $J -$ W2.  In Figure 8, the kink in the thick water cloud model sequence at $M_{\rm W2} \approx 16$  is due to the combination of changing opacity and pressure-temperature structure between 325 and 300~K, as water condenses into ice.

The  model trends are in reasonable agreement with the data, for example Figure 7 shows that the models reproduce the redward turn in $J - H$ at $M_J \sim 21$ or $T_{\rm eff} \sim 400$~K.  There are problematical discrepancies however, such as the divergence between models and data in $J - K$, for  $M_J > 21$. Variations in gravity cannot account for the discrepancies: increasing gravity improves the match for $J - H$ but makes it worse for $J - K$, for example. Note that all models are for solar metallicity, and it is likely that the sample contains brown dwarfs with a range of metallicity. This is one area that the models still need to address.

For the  $T_{\rm eff} \leq$ 450~K models, we find that increasing the cloud cover fraction (from 50\% to 80\%) and decreasing the sedimentation efficiency (from  $f_{\rm sed} = 5$ to  $f_{\rm sed} = 3$) greatly improves the agreement with  most of the observed colors. To explore the validity of this change, we compare the observed near-infrared spectrum of the Y0 dwarf W1217B (Leggett et al. 2014) to $T_{\rm eff} = 400$~K $\log g =$ 4.48 synthetic spectra in Figure 10. Here the observed flux has been scaled to 10~pc using the trigonometric parallax measured by Dupuy \& Kraus (2013), and the model fluxes are scaled to 10~pc adopting a brown dwarf radius of  0.1054 R/R$_{\odot}$ based on the evolutionary models of Saumon \& Morley (2008). Spectra for other Y0 dwarfs with trigonometric parallaxes are very similar in brightness, but have less complete wavelength coverage. We find that the 1.0 -- 1.3 $\mu$m spectrum in fact supports very thin to no cloud, and {\it not} the thick cloud model.

The large photometric discrepancies ($\sim 1$ magnitude) 
at $T_{\rm eff} \approx$ 400~K (spectral type Y0, $M_J \approx 20$, $M_{\rm W2} \approx 15$) consist of the models being too blue in $J - H$, $J - K$ and W2 $-$ W3, and too red in  [3.6] $-$ W2.  Adding more extensive and thicker water clouds appears to address the problems in $J - H$, $J - K$ and W2 $-$ W3 by reducing the flux at $J$, and increasing the flux at $K$ and W3.  However the spectral comparison shows that the apparent improvement is misleading --- in fact the $J$ flux becomes much too weak if the extent and thickness of the water clouds are increased. Instead, the near-infrared spectral comparison shown in Figure 10 and the  photometric comparisons shown in Figures 7 --- 9 indicate that the patchy cloud model fluxes for Y0 dwarfs are too low, by about a factor of two or a magnitude, at $Y$, $H$, $K$, [3.6] and W3. For later Y-types with  $T_{\rm eff} \approx$ 300~K, additional discrepancies seem to arise.

We discuss possible solutions to these problems in the next section. 

\section{Discussion}

\subsection{Y0 Dwarfs, $T_{\rm eff} \approx$ 400~K}

As noted above, for Y0 dwarfs with  $T_{\rm eff} \approx$ 400~K,
we have found that our state-of-the-art models generate fluxes that are  about a factor of two too low at  $Y$, $H$, $K$, [3.6] and W3 (filter profiles are shown in Figure 6).
For the $Y$, $H$ and W3 bandpasses  (the $\sim$ 1.05, 1.6 and 12.0 $\mu$m spectral regions)  
Figure 5 shows that  the dominant opacity source is NH$_3$. Figure 5 also includes cross sections for an arbitrary reduction in NH$_3$ abundance of a factor of two. In that case the dominant opacity in the $Y$-band and in the blue wing of the $H$-band (1.5 -- 1.6  $\mu$m) becomes pressure-induced H$_2$. The result would be that the pronounced double absorption feature at 1.03  $\mu$m goes away, 
and the flux in the $Y$ and $H$-bands significantly increases, all of which is in agreement with the observations (see Figure 10).
Note that the dominant opacity in the $J$-band is already  H$_2$, and the agreement between the cloud-free or thin-cloud models and the data are good for $J$ (Figure 10).

For most of the W3 band (8 -- 17 $\mu$m) the dominant opacity is NH$_3$ and remains so even if the abundance is halved. However the absorption would be dramatically weakened -- note that Figure 5 is plotted on a log scale. In this band, weakening the  NH$_3$ absorption could mimic the gain caused by the redistribution of flux due to increasing cloud cover (Figure 6).

With ammonia absorption significantly reduced, the remaining problems would be too little flux in the $K$- and [3.6]-bands. The 2 -- 4 $\mu$m flux is sensitive to CH$_4$ opacity (Figure 5) and new models should be calculated with a more complete  CH$_4$ linelist (although, as mentioned previously, at these very low temperatures large changes are not expected). The pressure-temperature structure of new models will in any case change  once the  change in the NH$_3$ absorption is incorporated, and this alone is likely to impact the  2 -- 4 $\mu$m flux.

For higher $T_{\rm eff}$ models than considered here, convective mixing leads to more N$_2$ and less NH$_3$  in the spectrum-forming part of the
atmosphere (e.g. at  $T_{\rm eff} \approx$ 600~K, Leggett et al. 2009). However Morley et al. (2014, their Figure 16) show that
preliminary models which include vertical mixing do not resolve the discrepancies for 400~K brown dwarfs --- the introduction of disequilibrium chemistry has a large impact on the $H- K$ color in particular, and makes the disagreement between models and data worse. These atmospheres are extremely complex, with low-lying chloride and sulphide clouds,  high water clouds and possibly multiple convective zones (Morley et al. 2014, their Figure 4). Including
vertical mixing in this context in a consistent fashion is challenging. The next stage in model generation will likely require an inversion approach, where a retrieval algorithm iterates both the pressure-temperature profile and the abundances to find the best match to the data (Line et al. 2014).

\subsection{Y1 Dwarfs, $T_{\rm eff} \approx$ 350~K}

For brown dwarfs with $M_{\rm J} \gtrsim 22$ and  $T_{\rm eff} \approx$ 350~K, additional issues with the models appear. The discrepancy at $J - H$ and $J - K$ increases by around 0.5 magnitudes (Figure 7). In the mid-infrared it appears that  $M_{\rm W2}$ increases more slowly with $J -$ W2 than calculated (Figure 8). These could point to a significant reduction of the 1 $\mu$m flux, of about a factor of two, so that  $J - H$, $J - K$  and  $J -$ W2
become much redder.  

At  $T_{\rm eff} \approx$ 350~K thick water clouds may be required to reproduce the data, as Figure 10 shows that such clouds cause a drop in the 1 $\mu$m flux of about the required size. Also the thick water cloud sequences in Figure 8 show interesting trends at $ 300 \lesssim T_{\rm eff}$~K$ \lesssim 350$ which are similar to those observed. The thick water cloud model has a shoulder in the trend of increasing $M_{\rm W2}$ with increasing  $J -$W2, followed by a sharp drop in  $M_{\rm W2}$. This flattening of the decrease in $M_{\rm W2}$ followed by a dramatic drop  could explain both the brightness of W1828 {\it and} the faintness of W0855, in W2 ($5 $ $\mu$m;  see \S 7.4).

Figure 11 reproduces the $J -$ W2:$M_{\rm W2}$ diagram. A simple solution to the apparent brightness of W1828 is that it is an equal-mass binary (see discussion in \S 7.4). Making this assumption, we have 
fit a polynomial to the observed  $J -$ W2:$M_{\rm W2}$ data, weighted by the inverse of the square of the uncertainty in the absolute magnitude, excluding known binaries. We find that the fifth-order polynomial
$$M_{W2} = 8.93387 + 4.37546*x - 1.80318*x^2 + 0.376357*x^3 - 0.0359539*x^4 + 0.00126562*x^5 $$
reproduces the trend seen in the observed data with W1828 as an equal-mass binary (Figure 11). 
The polynomial fit is not scientifically significant and we have not derived any measures of quality of fit.  We would expect the sample to have a range in metallicity, gravity and cloud properties which would result in scatter in the $J -$ W2:$M_{\rm W2}$ diagram (see \S 7.3). The function is useful, however, as a reference for future observations and theoretical modelling. The function does show similarity to the thick water cloud model sequence in Figure 8, which gives some supoprt to its validity. If that model sequence is shifted to fainter $J$  and brighter W2 magnitudes, say $J + 1$ and W2 $-$ 0.3, then there would be a shoulder at $M_{\rm W2} \sim 15.5$, followed by rapidly increasing W2 for $J - W2 \gtrsim 9$, as seen in Figure 11.
  
Water ice scatters at $J$ and absorbs at W2, and so these colors likely constrain  the ice particle size and composition, as well as cloud cover. 
Clearly this is a very interesting and challenging parameter space for atmospheric modeling.  
An inversion approach may also be necessary to improve the models at these temperatures.
Note that for  $T_{\rm eff} < 300$~K vertical transport no longer affects
the NH$_3$ abundance because it is far more abundant than N$_2$ down to deep
levels in the atmosphere.

\subsection{Intrinsic Scatter}

There is a large range, of about one magnitude,   in the observed $Y - J$ and $J - K$ colors for  $T_{\rm eff} \sim 400$~K (Figures 7 and 9).   The models in this temperature range show that $Y$, $J$ and $K$ are sensitive to the details of the cloud structure (Figures 6 and 10). Also, we have shown that the NH$_3$ absorption is likely to be much lower than calculated, implying that pressure-induced H$_2$ becomes an important opacity source at 1.55$\,\mu$m and dominant at 1.05, 1.25 and 2.05$\,\mu$m (Figure 5). These wavelength regions coincide with the $YJHK$ filter bandpasses which were designed to avoid the Earth's   H$_2$O absorption bands. The observed near-infrared colors will become a function of the H$_2$ absorption, which in turn is dependent on metallicity and gravity (Saumon et al. 2012). The spread in the  near-infrared colors for spectral types around Y0 therefore likely reflects  a range of  metallicity, gravity and cloud properties.

\subsection{WISEP J182831.08$+$265037.8 and  WISE J085510.83$-$071442.5}

Of the known Y dwarfs, the reddest in $J -$W2 are  W1828 (Y1.5) and W0855 (nominally Y2). These are likely to be the coolest of the known Y dwarfs.

W1828 appears unusually bright in the mid-infrared, and has been widely discussed in the literature (Beichman et al. 2013 and references therein). The WISE atlas image does not show any obvious infrared cirrus coincident with the source, which might lead to a mid-infrared excess. This work and other recent publications 
have produced more photometry and parallaxes for comparable objects, and we can explore the nature of W1828 further. 

We have calculated the colors of a binary system composed of  W0647 (Y1) and  W0855.  A simple composite of these two sources put at the same distance does not produce a binary  that is sufficiently red in $J -$W2 however. This is because W0855 is intrinsically faint and does not contribute enough to the mid-infrared  total flux. Instead, as mentioned in \S 7.2,  we suggest that W1828 is a simple pair of identical brown dwarfs. Each component would have  $M_J \approx 24.4$ and $M_{\rm W2} \approx 15.2$. Such a source has colors consistent with the trends seen in Figures 7 to 9.   Binaries are not uncommon in brown dwarf samples --- Radigan et al. (2013) find a volume-corrected binary fraction for L9 to T4 types of $13^{+7}_{-6}$\%, similar to values reported for other brown dwarf spectral types. Also, Burgasser et al. (2007) find that very low mass multiple systems tend to be closely separated and are more frequently in near-equal mass configurations.

The binary explanation of W1828 would imply  that it consists of a pair of $ 300 \lesssim T_{\rm eff}$~K$ \lesssim 350$ brown dwarfs. 
For W0855, our models together with the observed $J$- and W2-band luminosity  imply that it is a $\sim$250~K dwarf. If 
W1828 and W0855 are say 3 Gyrs old (being in the solar neighbourhood), evolutionary models show that their masses are $\sim$10 and $\sim$5 Jupiter-masses respectively  (Saumon \& Marley 2008).

\section{Conclusion}

We have found that current  state-of-the-art models can qualitatively reproduce  the trends seen in the observed colors of the latest-type T dwarfs and known Y dwarfs. However significant discrepancies exist. For brown dwarfs with $T_{\rm eff} \approx$ 400~K, corresponding to spectral type $\sim$Y0, the model fluxes are  a factor of two low at $Y$, $H$, $K$, [3.6]  and W3. For  $T_{\rm eff} \approx$ 350~K, corresponding to spectral type $\sim$Y1, it appears that a reduction in the $J$-band flux of about a factor of two is needed, while the W2 flux remains approximately constant.

The problems at $Y$, $H$ and W3 (or $\sim$ 1.05, 1.6 and 12.0 $\mu$m) may be addressed by significantly reducing the NH$_3$ absorption, for example by halving the abundance of NH$_3$. This could be explained by mixing of N$_2$ and  NH$_3$, as is seen in T dwarfs, but incorporating mixing in a realistic way in an atmosphere with chloride, sulfide and water condensates is challenging.  Once the NH$_3$ absorption is reduced, pressure-induced H$_2$ becomes the dominant opacity source in the near-infrared filter bandpasses, making those colors sensitive to gravity and metallicity variations, and probably explaining the scatter seen in the near-infrared colors of Y0 dwarfs. 

While the cloud layers are thin to non-existent for the 400~K brown dwarfs, as $T_{\rm eff}$ decreases to 350~K thick and extensive water clouds appear to form, based on the observed reduction in the  $J$-band flux. The onset of these clouds might occur over a narrow range of $T_{\rm eff}$, as indicated by the observed small change in 5 $\mu$m flux over a large change in $J - W2$ color. 

Diagnosis of cloud cover for Y dwarfs however will remain uncertain until the models, which are already complex, can be further improved to the point of reproducing the near- and mid-infrared observations. The expectation had been that the atmospheres of the cold brown dwarfs would be simpler, because most elements are condensed out and the chemistry is reduced to CH$_4$, H$_2$O and NH$_3$,  but this was clearly rather optimistic.

\acknowledgments

DS is supported by NASA Origins NNH12AT89I.
Based on observations obtained at the Gemini Observatory, which is operated by the Association of Universities for
 Research in Astronomy, Inc., under a cooperative agreement with the
    NSF on behalf of the Gemini partnership: the National Science
    Foundation (United States), the Science and Technology Facilities
    Council (United Kingdom), the National Research Council (Canada),
    CONICYT (Chile), the Australian Research Council (Australia),
    Minist\'{e}rio da Ci\^{e}ncia, Tecnologia e Inova\c{c}\~{a}o (Brazil)
    and Ministerio de Ciencia, Tecnolog\'{i}a e Innovaci\'{o}n Productiva
    (Argentina). SKL's research is supported by Gemini Observatory.  This publication makes use of data products from the Wide-field Infrared Survey Explorer, which is a joint project of the University of California, Los Angeles, and the Jet Propulsion Laboratory/California Institute of Technology, funded by the National Aeronautics and Space Administration. This research has made use of the NASA/ IPAC Infrared Science Archive, which is operated by the Jet Propulsion Laboratory, California Institute of Technology, under contract with the National Aeronautics and Space Administration.
This research has made use of the Keck Observatory Archive (KOA), which is operated by the W. M. Keck Observatory and the NASA Exoplanet Science Institute (NExScI), under contract with the National Aeronautics and Space Administration.



\clearpage

\begin{figure}
      \includegraphics[angle=0,width=0.8\textwidth]{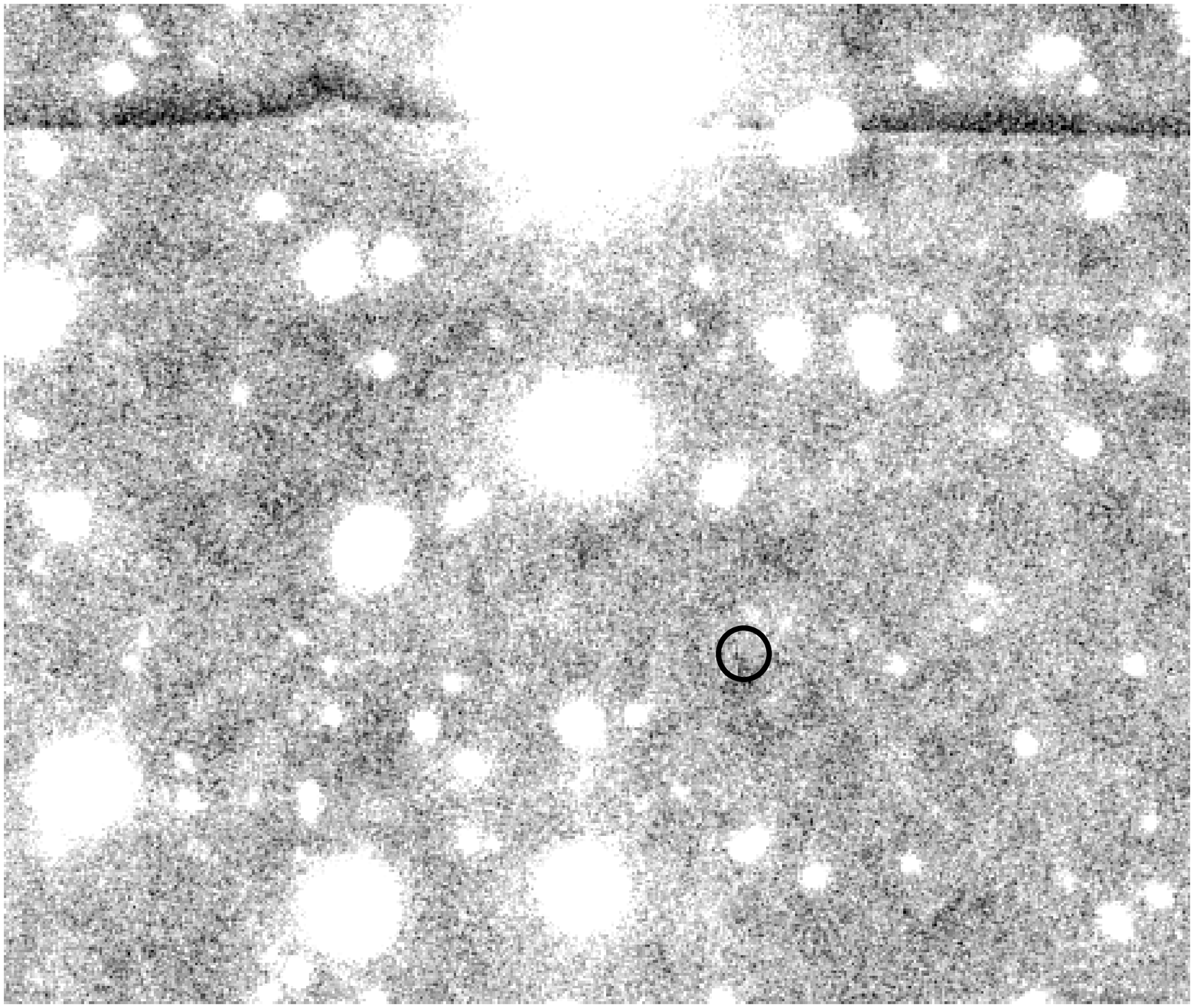}\\
  \includegraphics[angle=0,width=.4\textwidth]{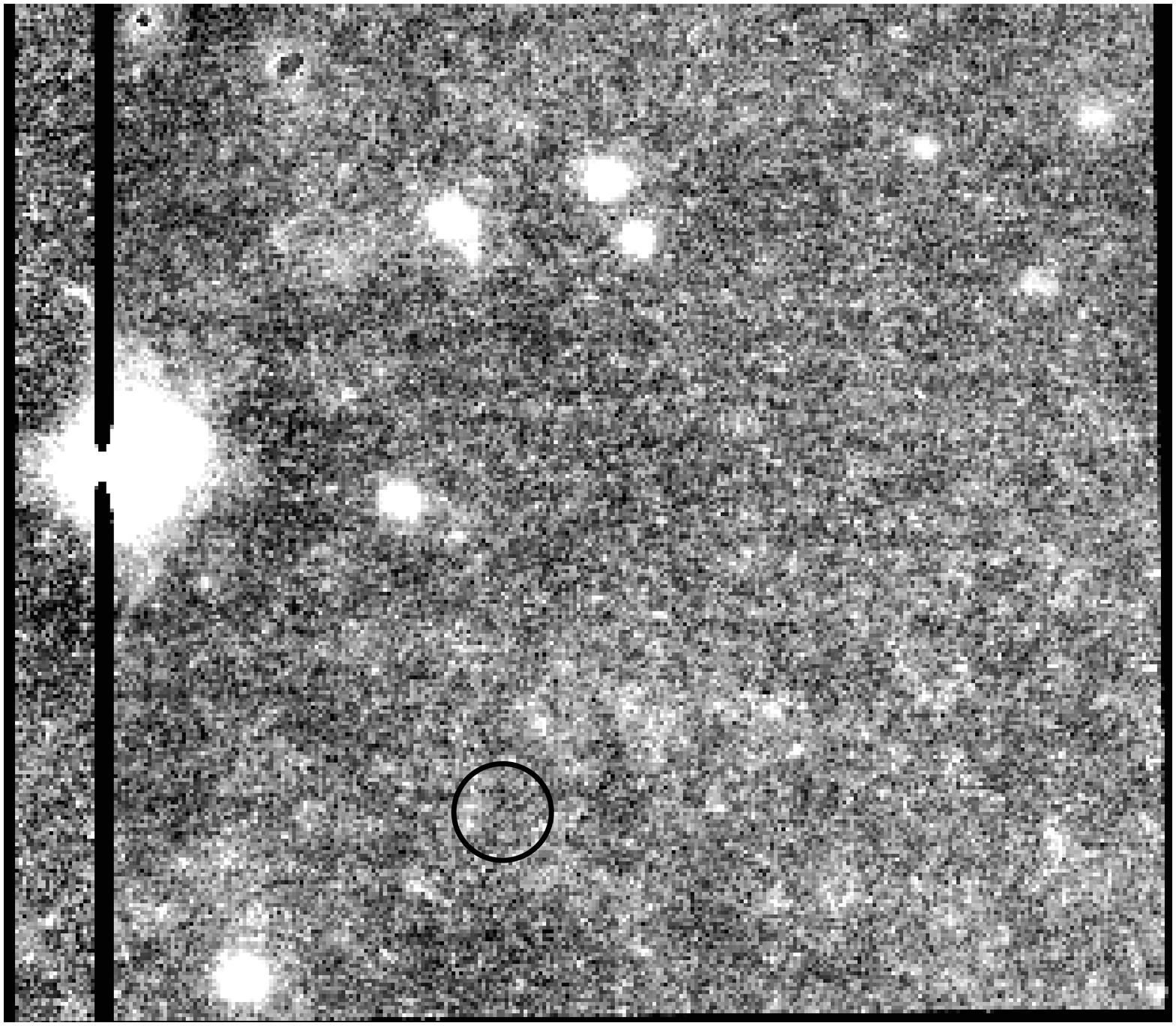}\hfill
 \includegraphics[angle=0,width=.4\textwidth]{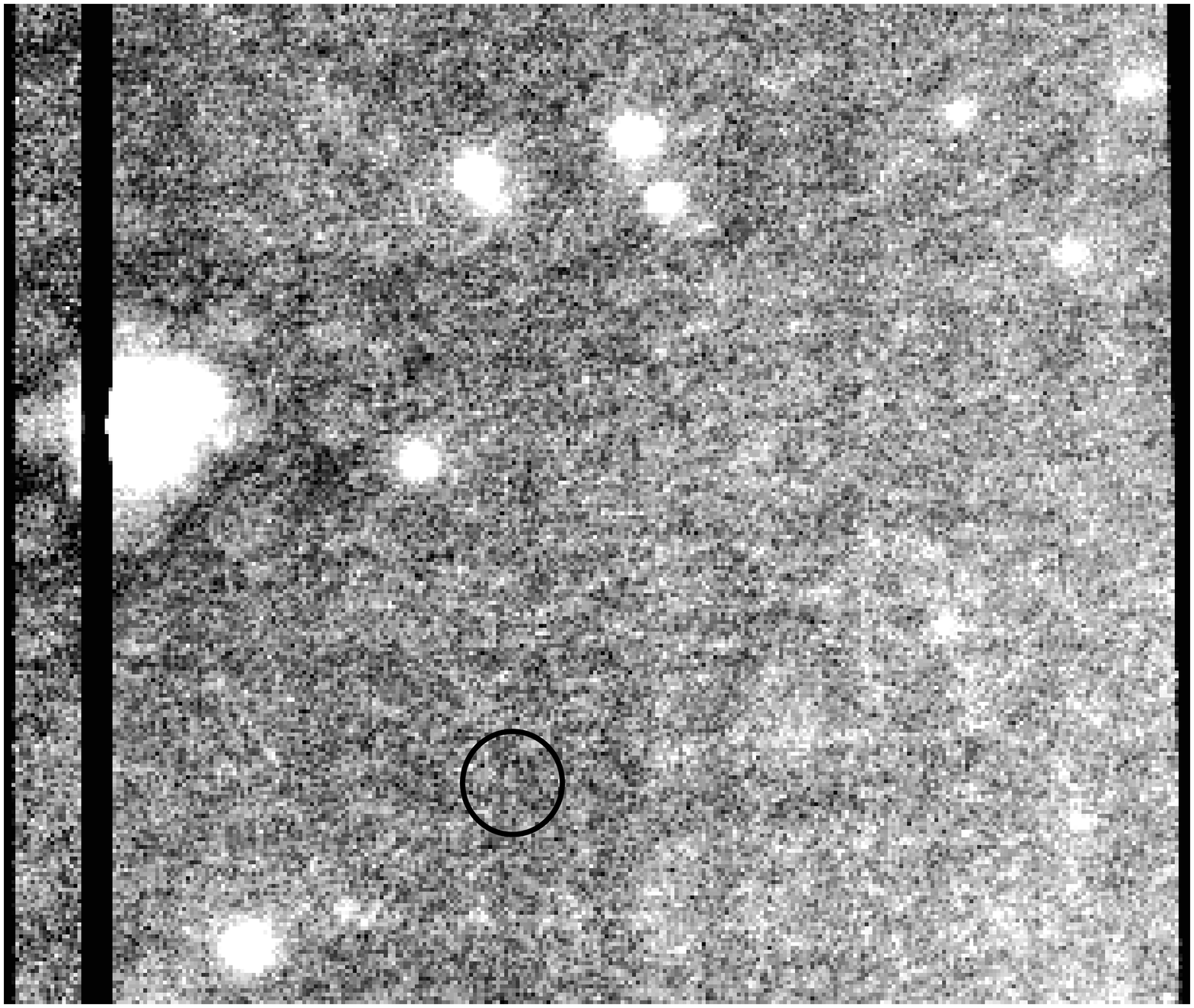}\\
\caption{Images of the field around WD  0806$-$66B. The top panel  is our coadded GMOS $z$ image, with North up and East to the left, approximately 1 arcminute on a side. The lower two panels are our GSAOI images,   $Z$-band on left (equivalent to MKO $Y$) and  $J$-band on right.  The images are approximately 25 arcseconds on a side. The expected location of the brown dwarf is shown as a black circle.
\label{fig1}}
\end{figure}

\begin{figure}
    \includegraphics[angle=0,width=.8\textwidth]{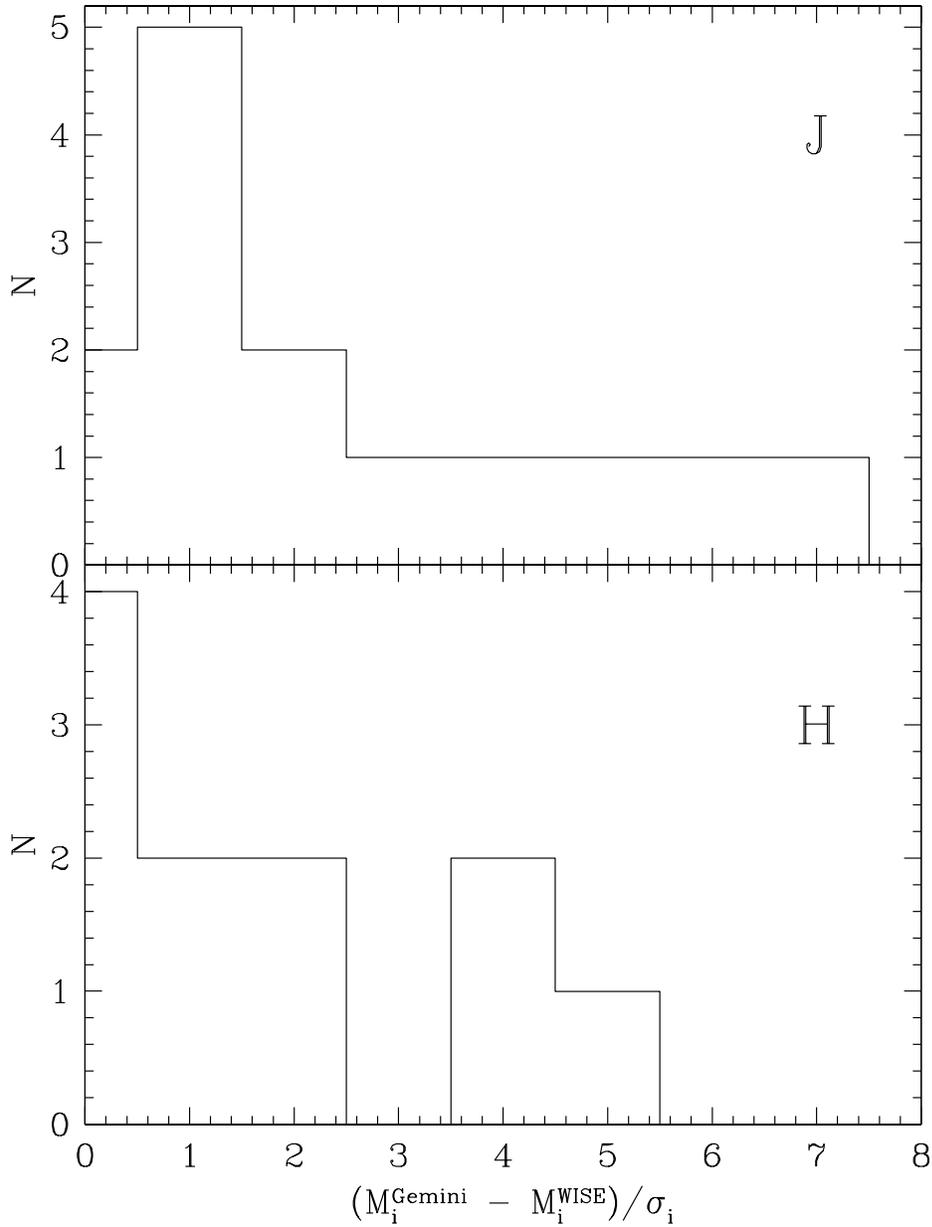}
\caption{A comparison of the $J$ and $H$ magnitudes published by our group (this work and Leggett et al. 2013) and the WISE group   (Beichman et al. 2013, 2014; Kirkpatrick et al. 2012, 2013; Mace et al. 2013). The bins  show the difference between the measured values from each group divided by the square root of the sum of the squares of the uncertainties quoted for each measurement. About one-third of the sample have measurements that differ by more than $2 \sigma$. 
\label{fig2}}
\end{figure}

\begin{figure}
    \includegraphics[angle=0,width=.85\textwidth]{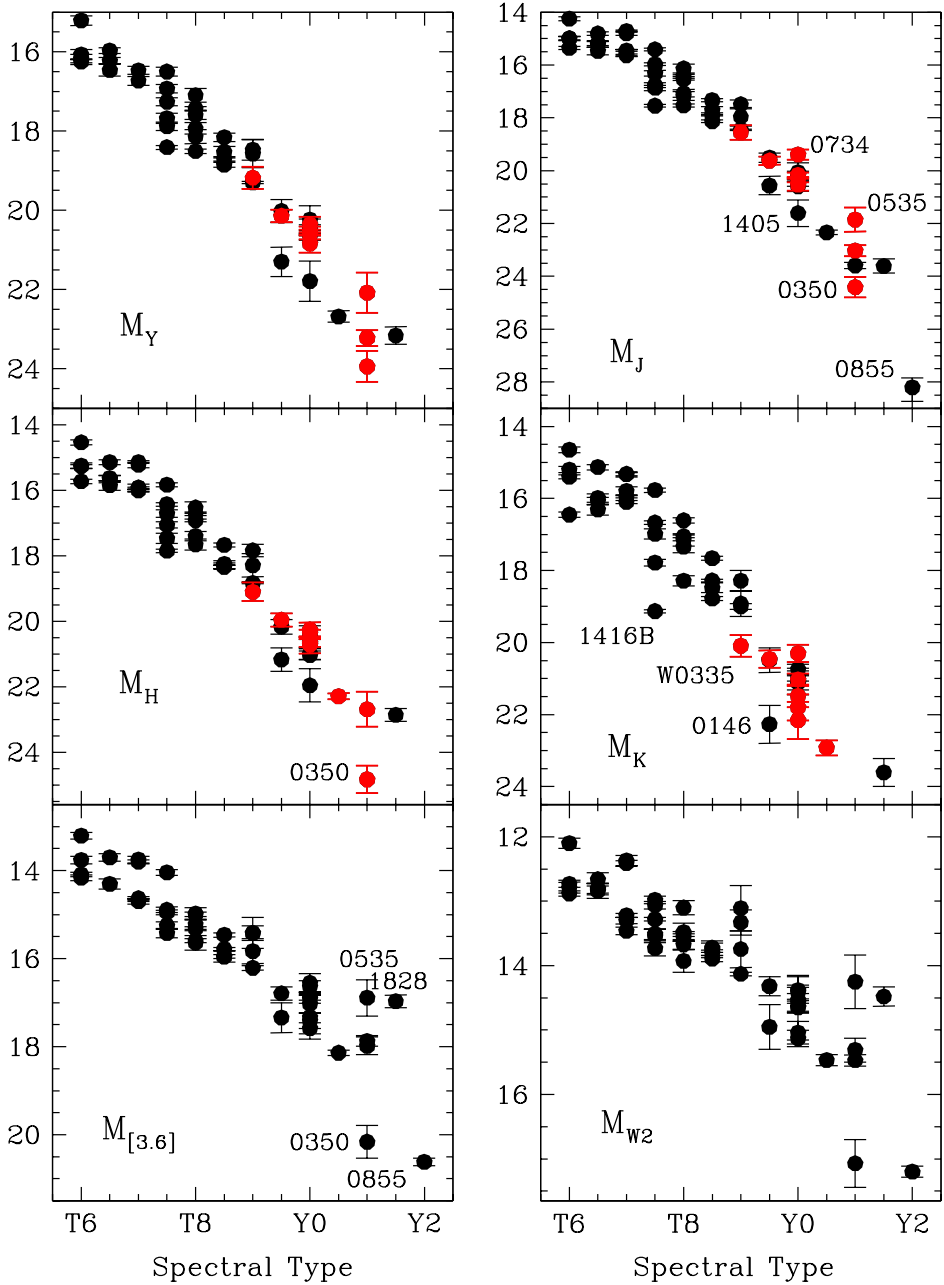}
\caption{Trends in absolute magnitude as a function of spectral type. Note the varied y-axis ranges. All magnitudes are Vega based. The $YJHK$ are on the MKO system, [3.6] on the {\it Spitzer} IRAC system and W2 (4.6 $\mu$m) on the {\it WISE} system. Photometry and parallax sources are given in the text. New observations presented here are identified by red symbols.
\label{fig3}}
\end{figure}

\begin{figure}
    \includegraphics[angle=0,width=.85\textwidth]{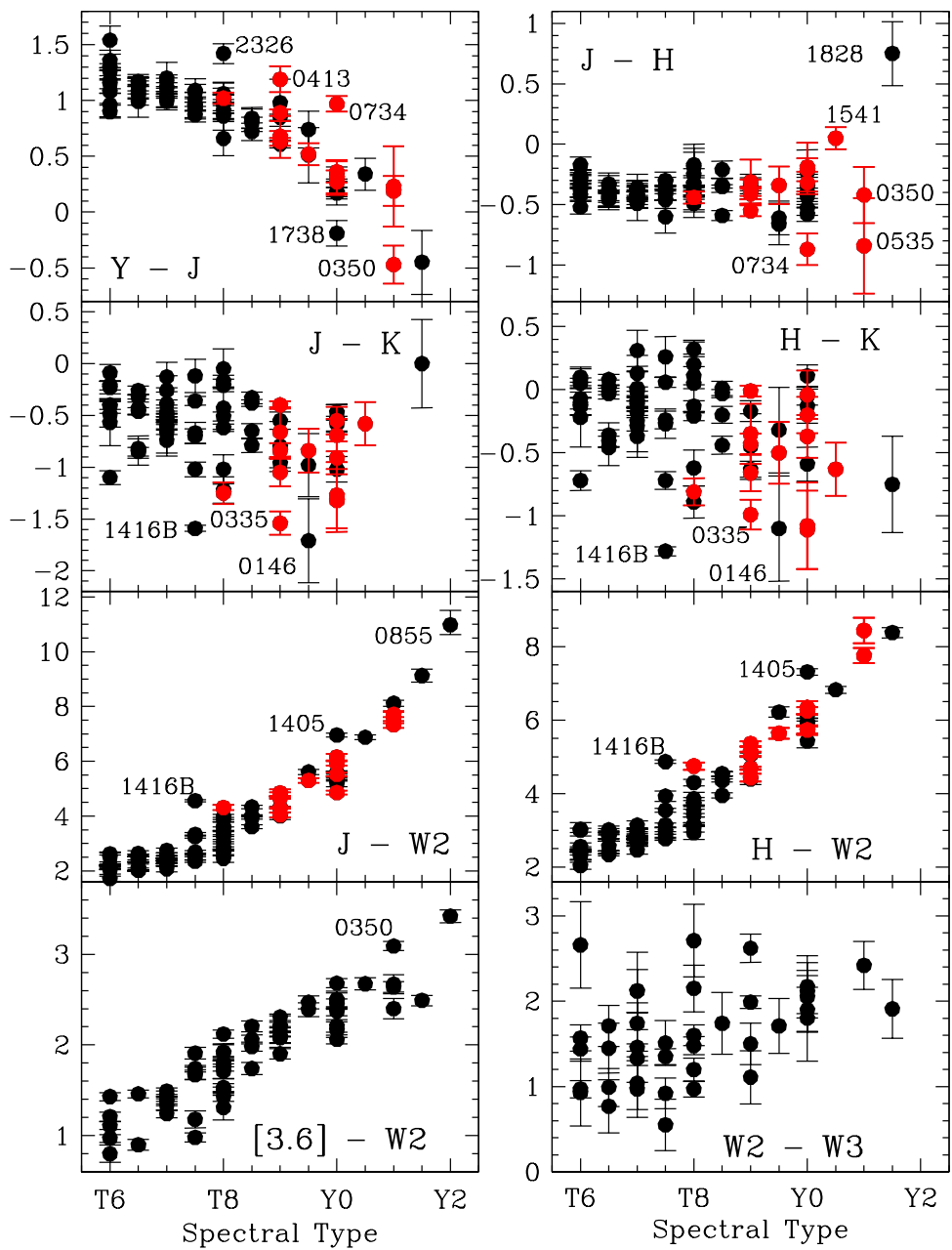}
\caption{Trends in color as a function of type. Data sources are given in the text. All magnitudes are Vega based. New observations presented here are identified by red symbols.
\label{fig4}}
\end{figure}

\begin{figure}
 \includegraphics[angle=0,width=1.1\textwidth]{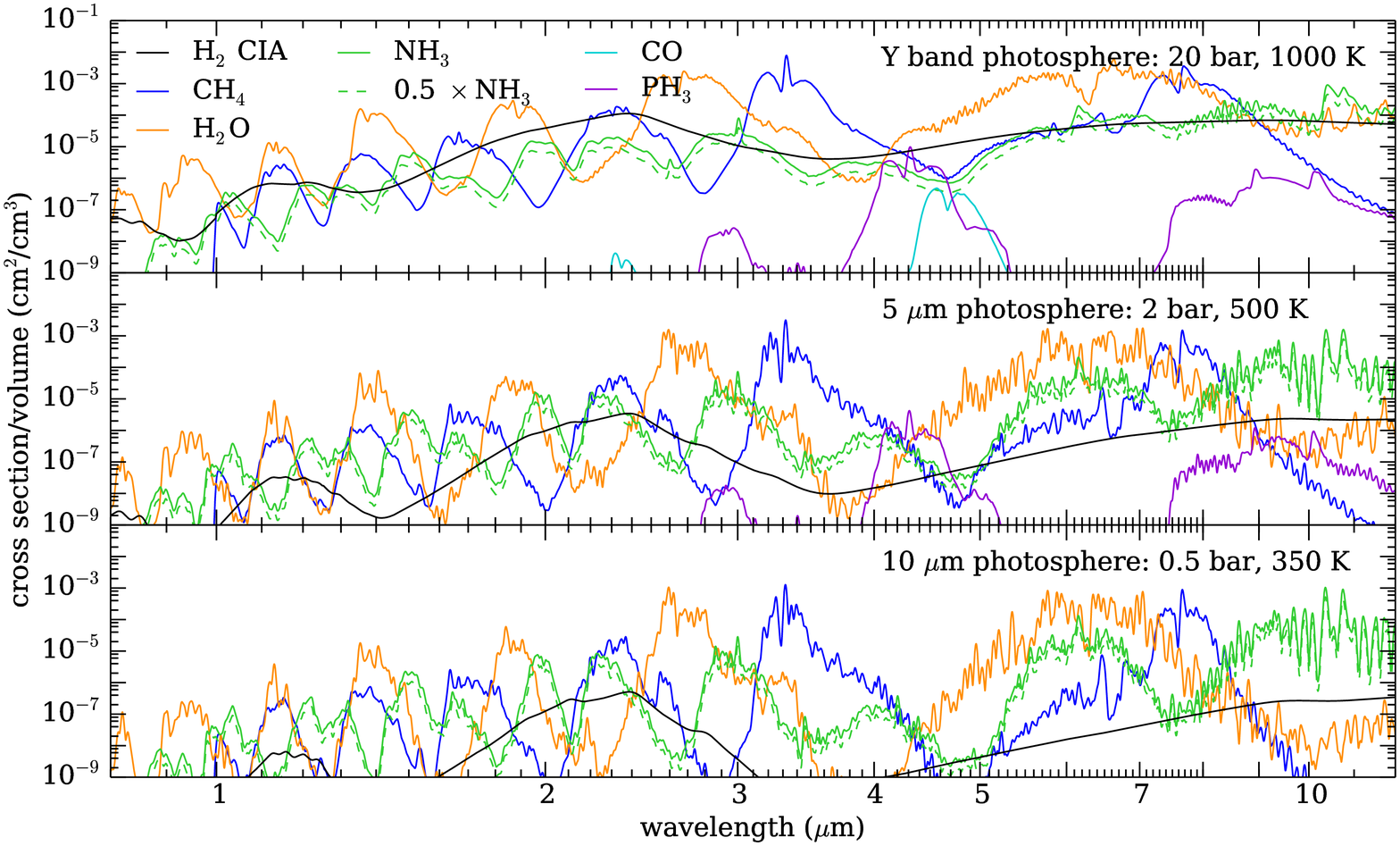}
\caption{Opacity cross sections  
multiplied by equilibrium gas volume mixing ratio, 
calculated for three layers of  a   $T_{\rm eff} = 400$~K, $\log g = 4.48$, brown dwarf atmosphere. The top panel shows the layer from which the $Y$-band flux emerges, the middle panel the 5 $\mu$m flux, and the bottom panel the  10 $\mu$m  flux. See text for discussion.
\label{fig5}}
\end{figure}

\begin{figure}
    \includegraphics[angle=0,width=0.7\textwidth]{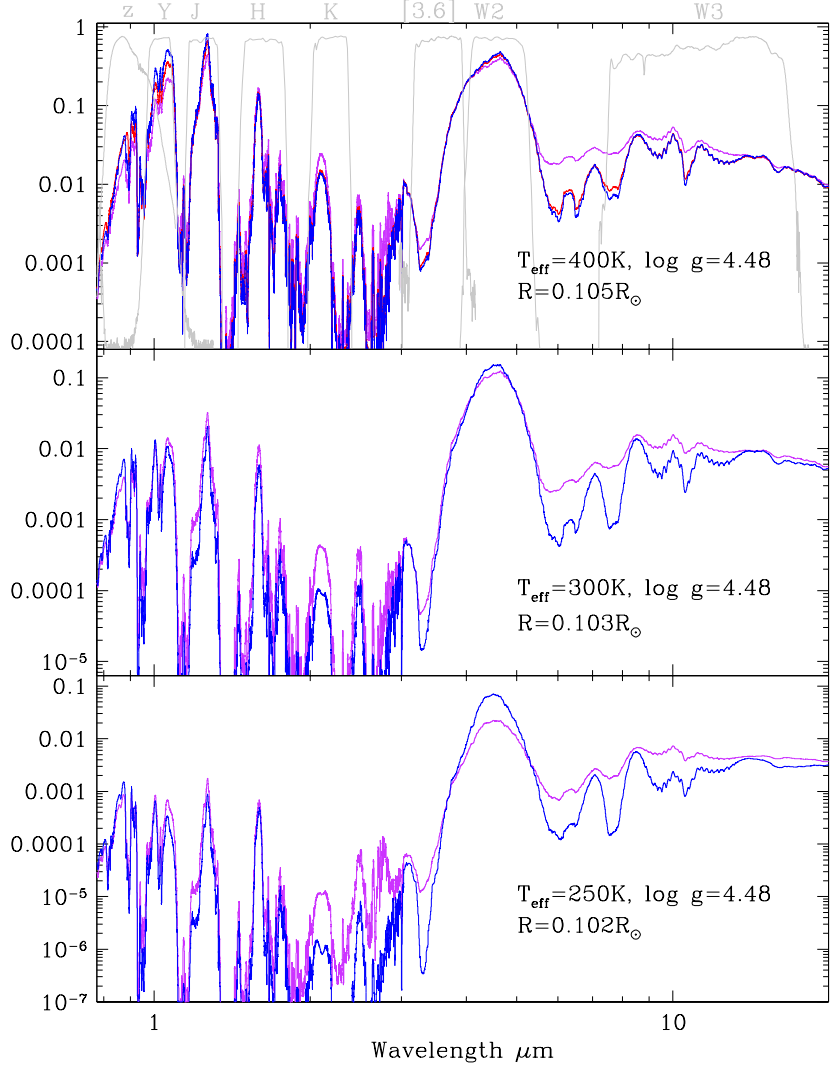}
\caption{Blue, violet and red lines are synthetic spectra for a   $\log g = 4.48$ brown dwarf at 10~pc. 
Spectra for three values of $T_{\rm eff}$ are shown, as indicated in the legend. The radii given by evolutionary models (Saumon \& Marley 2008), and used to scale the model fluxes, are also given in the legend. 
Flux is plotted as $F_{\lambda}$ in units of $10^{-16}$ W m$^{-2} \mu$m$^{-1}$. 
The red line has thin water cloud decks with $f_{\rm sed} = 5$ over half the surface, and the violet line has thick water cloud decks with $f_{\rm sed} = 3$ over 80\% of the surface. The blue line is for a cloud-free atmosphere. Gray lines in the top panel show photometric filter bandpasses as identified by the legend: $z$, $Y$, $J$, $H$, $K$,  {\it Spitzer} [3.6],  and  {\it WISE} W2 (4.6 $\mu$m) and W3 (12  $\mu$m).
\label{fig6}}
\end{figure}

\begin{figure}
    \includegraphics[angle=0,width=.7\textwidth]{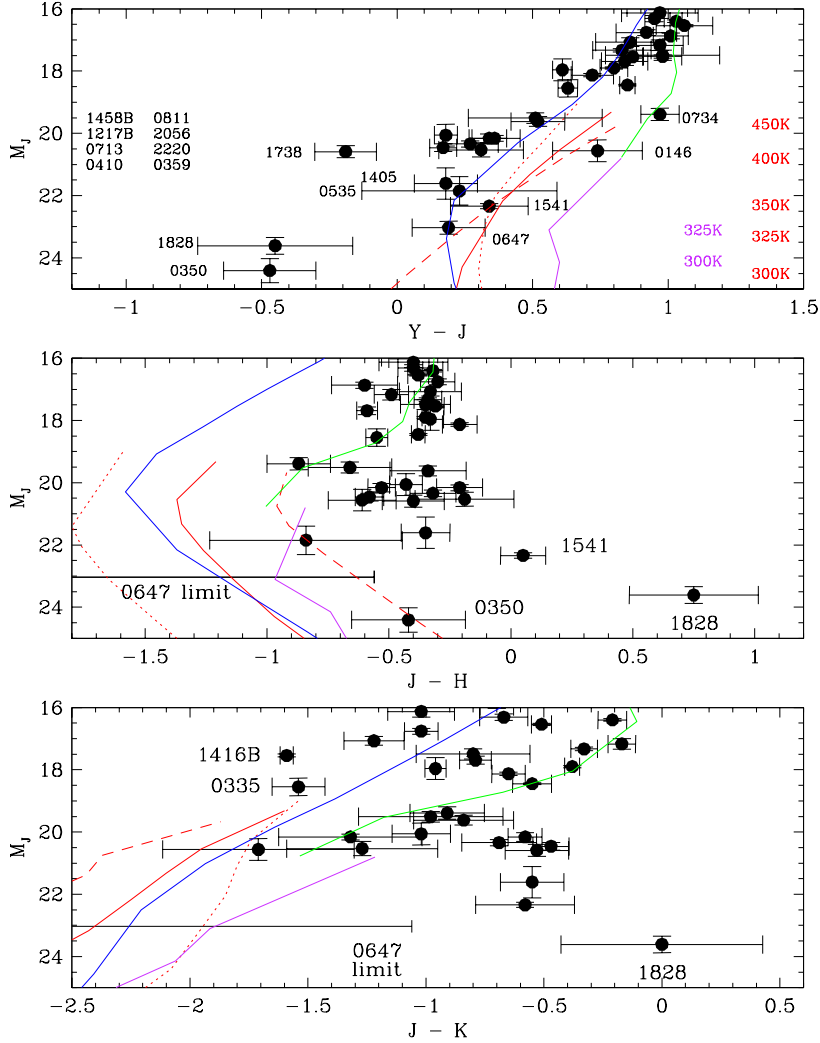}
\caption{Absolute $J$ magnitude as a function of near-infrared colors for T8 and later type dwarfs. 
Not shown are WD  0806$-$66B with $M_J = 23.6$, $z_{\rm AB} - J > 1.2$ and  $Y - J > -1.5$, and
WISE J085510.83$-$071442.5 with  $M_J \approx 28.3$ (see \S 2). WISE J035000.32$-$565830.2
appears unusually faint in $M_J$ (and $M_{\rm W2}$) and the parallax should be confirmed.
Model sequences (Saumon et al. 2012, Morley et al. 2012, 2014) are shown for $\log g = 4.48$: blue lines are 
cloud-free, green have thin sulfide and salt cloud decks with $f_{\rm sed} = 5$ over the entire surface, red
have  thin water cloud decks with $f_{\rm sed} = 5$ over half the surface, and violet 
have thick water cloud decks with $f_{\rm sed} = 3$ over 80\% of the surface.
The dashed line is for  $\log g = 5.0$ and dotted  for $\log g = 4.0$. The right axis of the top panel shows $T_{\rm eff}$ values for the  $\log g = 4.48$ water cloud models.
\label{fig7}}
\end{figure}

\begin{figure}
    \includegraphics[angle=0,width=.8\textwidth]{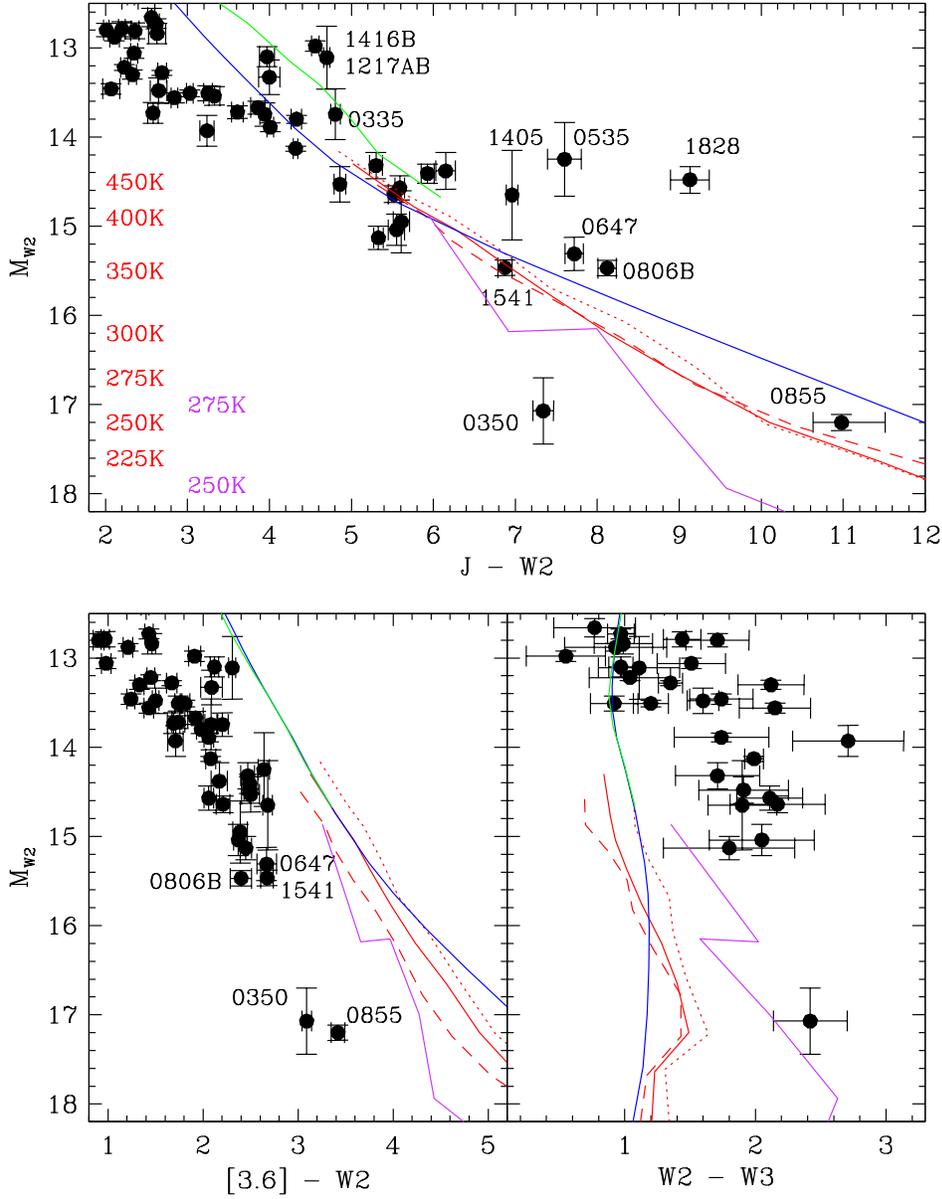}
\caption{Absolute W2 magnitude as a function of various colors for T7  and later type dwarfs.  For WD  0806$-$66B the W2 magnitude is uncertain ($\delta M = 0.4$) and we have replaced it with the {\it Spitzer} 4.48~$\mu$m magnitude transformed to W2 by adding 0.1 magnitude (Leggett et al. 2013). See note regarding WISE J035000.32$-$565830.2
 in the caption to Figure 7. Sequences are as in Figure 7. 
\label{fig8}}
\end{figure}

\begin{figure}
    \includegraphics[angle=0,width=.8\textwidth]{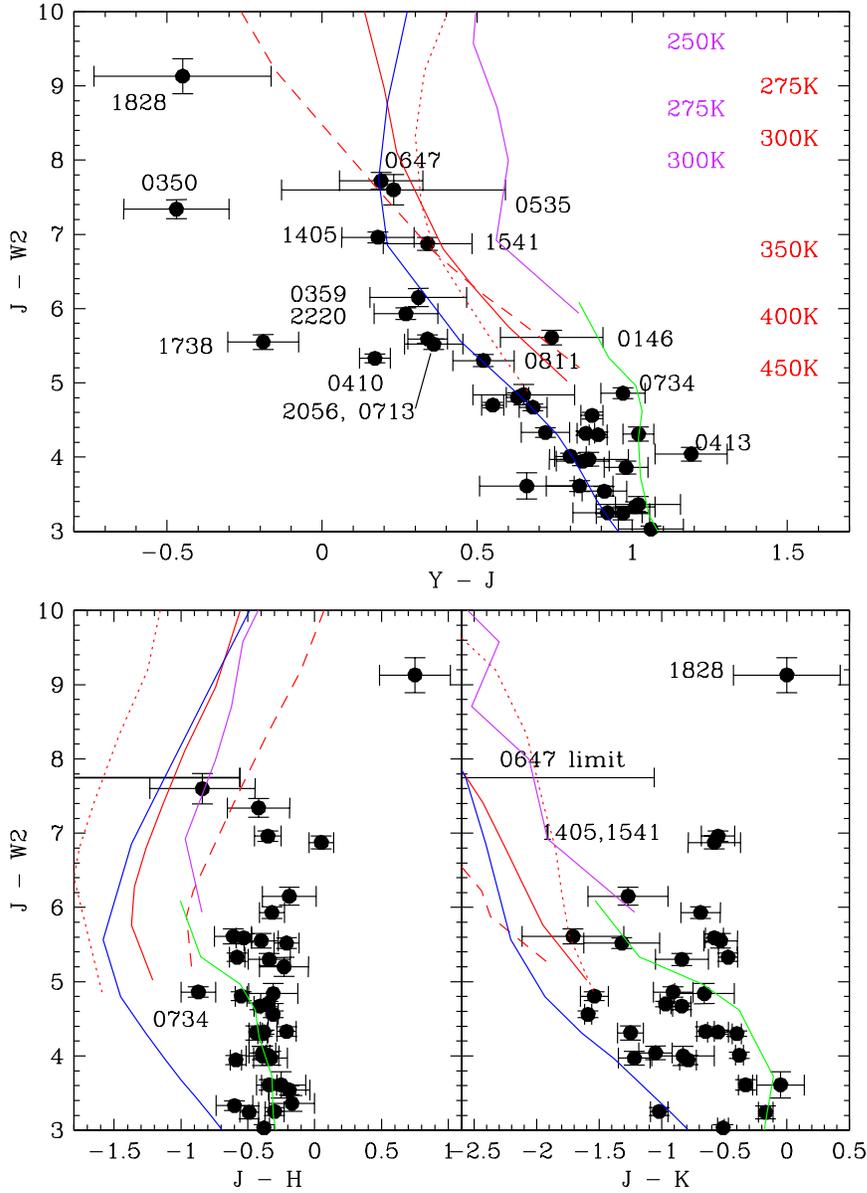}
\caption{$J -$ W2 as a function of color.  Not shown are WD  0806$-$66B ($J - W2 = 8.1$) and W0855 ($J - W2 \approx 11.0$), neither of which have measured $Y$, $H$ or $K$ values available. Sequences are as in Figure 7. \label{fig9}}
\end{figure}

\begin{figure}
    \includegraphics[angle=-90,width=1\textwidth]{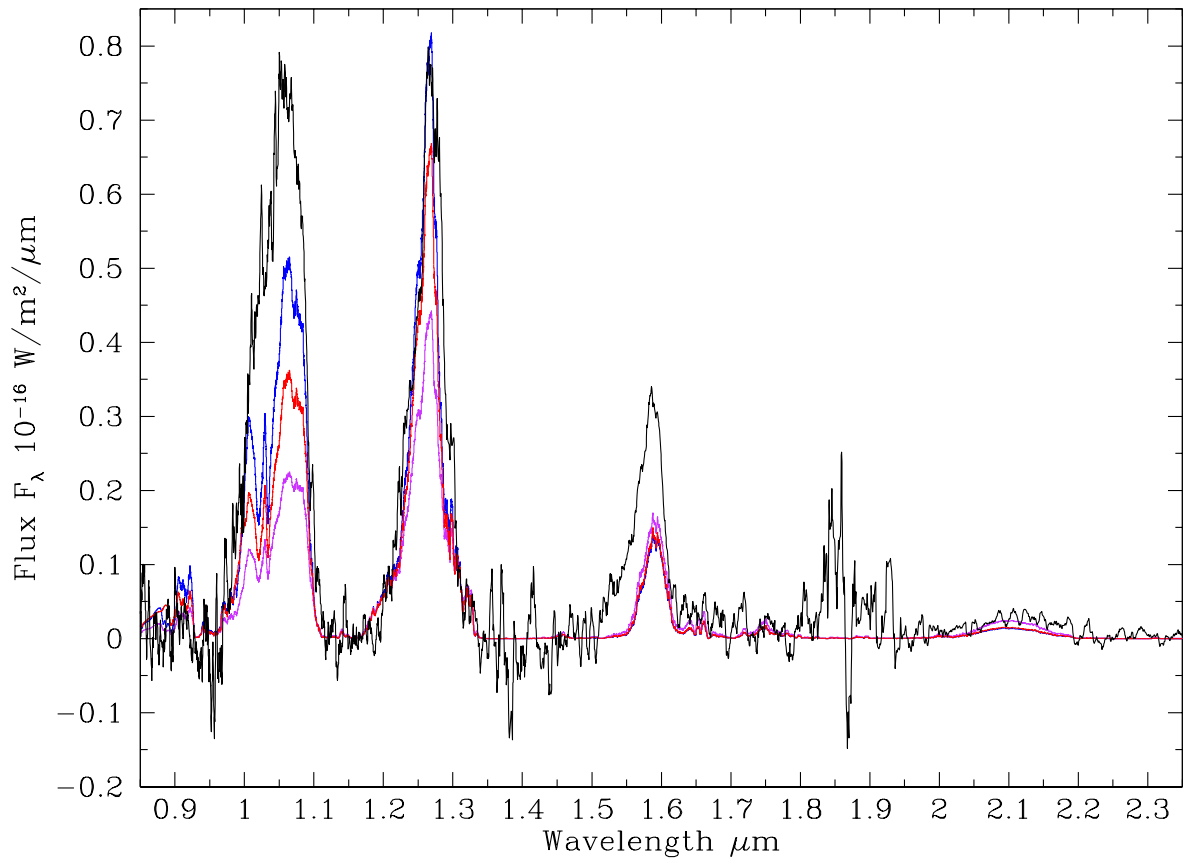}
\caption{The observed (smoothed) near-infrared spectrum for the Y0  WISEPC J121756.91$+$162640.2B (Leggett et al. 2014) is shown as a black line. The flux has been scaled to what would be observed were the dwarf at a distance of 10 pc. The region of poor atmospheric transmission 1.80 -- 1.94 $\mu$m is noisy. The blue, red and violet lines are synthetic spectra for a  $T_{\rm eff} = 400$~K $\log g = 4.48$ brown dwarf at 10~pc, with different cloud cover parameters as in Figure 7. \label{fig10}}
\end{figure}

\begin{figure}
    \includegraphics[angle=-90,width=1\textwidth]{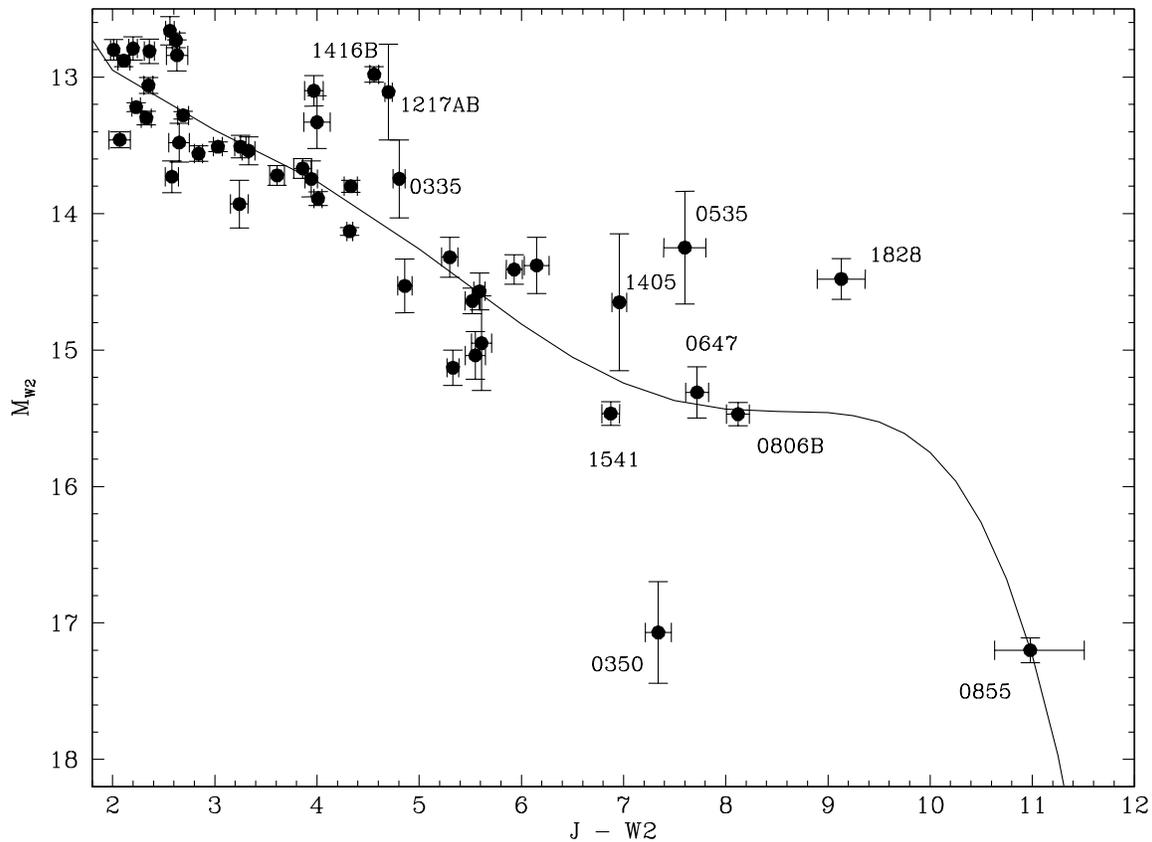}
\caption{The black line is fifth-order polynomial weighted fit to the {$J -$ W2, $M_{\rm W2}$} data points, excluding known binaries. The fit is not scientifically significant, and is intended only as an indication of observed trends, with the assumption that W1828 is an equal-mass binary. W0350 is included in the fit, however the parallax for this object appears to be erroneous. While  admittedly arbitrary, we note that a very similar trend can be obtained by shifting the thick water cloud model sequence in $J -$ W2 and $M_{\rm W2}$ (Figure 8, violet line). 
\label{fig11}}
\end{figure}






\clearpage

\begin{deluxetable}{lcllllccr}
\tabletypesize{\scriptsize}
\tablewidth{0pt}
\rotate
\tablecaption{New $YJHK$ MKO Photometry}
\tablehead{
\colhead{Name} & \colhead{Spectral}  & \colhead{$Y$(err)\tablenotemark{a}} & \colhead{$J$(err)} & 
\colhead{$H$(err)} & \colhead{$K$(err)\tablenotemark{a}} & \colhead{Date}  & \colhead{Instrument} &  \colhead{Discovery}\\
\colhead{} & \colhead{Type} & \colhead{} & \colhead{} & \colhead{} & \colhead{}  
& \colhead{YYYYMMDD} & \colhead{}  & \colhead{Reference}\\
}
\startdata
WISE J000517.48$+$373720.5 & T9 &   18.48(0.02) &  17.59(0.02) &  17.98(0.02) &  17.99(0.02) & 2013 0717 & NIRI & M13 \\
WISE J001354.39+063448.2 & T8 &    20.56(0.04) & 19.54(0.03) & 19.98(0.04) & 20.79(0.10) & 2013 0812& NIRI & P14a \\
WISE J033515.01$+$431045.1 & T9 &     19.95(0.03) & 19.32(0.02) & 19.87(0.04) & 20.86(0.11) & 2013 1127 & NIRI & K12 \\
WISE J035000.32$-$565830.2 & Y1 &    21.62(0.12) &  22.09(0.12) & 22.51(0.20)   &   \nodata  &  2014 0104,0108,1008,1104 & FLAMINGOS-2 & K12 \\
WISE J035934.06$-$540154.6 & Y0 &    21.84(0.11) & 21.53(0.11) & 21.72(0.17) & 22.8(0.3) & 2013 1121,1122 & FLAMINGOS-2 & K12 \\
WISE J041358.14$-$475039.3 & T9 &     20.82(0.10) &  19.63(0.06) &  20.02(0.08) & 20.68(0.12) &  2013 1121 & FLAMINGOS-2 & M13 \\
WISE J053516.80$-$750024.9 & Y1   &  22.73(0.30) & 22.50(0.20)  & 23.34(0.34)   & \nodata  & 2014 0105,0106,1008 & FLAMINGOS-2 & K12 \\
WISE J064723.23$-$623235.5\tablenotemark{b} & Y1  &   23.13(0.09) &  22.94(0.10) & $>23.5$ & $>24$ & 2014 0111,0220,0222,0223  & FLAMINGOS-2 & K13\\
WISE J071322.55$-$291751.9 & Y0  &   20.34(0.08)  & 19.98(0.05) & 20.19(0.08) & 21.30(0.31) &   2013 0112 & NIRI & K12 \\
WISE J073444.02$-$715744.0 & Y0  &   21.02(0.05) &  20.05(0.05) & 20.92(0.12) & 20.96(0.15) & 2013 1127 & FLAMINGOS-2 &  K12 \\ 
WISEPA J075108.79$-$763449.6 & T9 &    20.02(0.10) & 19.37(0.13) & 19.68(0.13) &  20.03(0.20) & 2013 1129 & FLAMINGOS-2 &  K11 \\
WD 0806$-$66B\tablenotemark{b,c} & Y1  & $> 23.5$ & $> 23.9$ &  \nodata & \nodata & 2013 1217,1218; 2014 0115,0116,0119 & GSAOI &  L11\\
WISE J081117.81$-$805141.3 & T9.5  &   20.17(0.07) &  19.65(0.07) &  19.99(0.14) & 20.49(0.20) & 2013 1126,1129,1206 &  FLAMINGOS-2 &  K12 \\
WISEPC J140518.40$+$553421.5 & Y0  &    \nodata  &  \nodata &  \nodata &  21.61(0.12) & 2013 0421 & NIRI & C11 \\
WISEP J154151.65$-$225025.2 & Y0.5  &    \nodata  &  \nodata &  21.07(0.07) & 21.7(0.2) &  2013 0508,0511,0805 & NIRI & C11 \\
WISEPA J161441.45$+$173936.7 & T9  &   19.58(0.04) & 18.90(0.02) & 19.31(0.04) & 19.74(0.07) & 2013 0404 & NIRI & K11 \\
WISE J222055.31$-$362817.4 & Y0  &   20.91(0.09) & 20.64(0.05) & 20.96(0.08) & 21.33(0.15) & 2014 0716 & NIRI & K12 \\
\enddata
\tablenotetext{a}{The NIRI $Y$ magnitudes and FLAMINGOS-2 $Ks$ have been put on the MKO $Y$ and $K$ system as described in the text.}
\tablenotetext{b}{Where given, lower limit magnitudes correspond to the limit for a $3 \sigma$ detection.}
\tablenotetext{c}{For WD 0806$-$66B a limit of $z_{\rm AB} > 26.2$ was obtained using GMOS-South on 2011 1205, 1230, 1231 and 2012 0119, 0120 and 0121.}
\tablecomments{Discovery references are: Cushing et al. 2011; Kirkpatrick et al. 2011, 2012 and 2013; Luhman et al 2011; Mace et al. 2013; Pinfield et al. 2014a.}
\end{deluxetable}

\clearpage

\begin{deluxetable}{lcrllllccc}
\tabletypesize{\scriptsize}
\tablewidth{0pt}
\rotate
\tablecaption{MKO Photometry of WISE Y Dwarfs}
\tablehead{ 
\colhead{Name} & \colhead{Spectral}  & \colhead{$M - m$(err)} & \colhead{$Y$(err)} & \colhead{$J$(err)} & \colhead{$H$(err)} & \colhead{$K$(err)} & \colhead{Parallax} & \colhead{Photometry} & \colhead{Discovery}\\
\colhead{}     & \colhead{Type}      & \colhead{magnitude}    & \colhead{}         & \colhead{}         & \colhead{}         & \colhead{}         & \colhead{Reference}  & \colhead{Reference} & \colhead{Reference}\\
}
\startdata
WISE J030449.03$-$270508.3 & Y0  & \nodata    & \nodata    & 20.79(0.09) & 21.02(0.16)  & \nodata     & \nodata    & P14b & P14b \\
WISE J035000.32$-$565830.2   & Y1   & 2.32(0.37) & 21.62(0.12) & 22.09(0.12) & 22.51(0.20)    &  \nodata    &  M13 & this work & K12 \\
WISE J035934.06$-$540154.6   & Y0   & $-$1.00(0.20)) & 21.84(0.11) & 21.53(0.11) & 21.72(0.17) & 22.8(0.3)   &  T14 & this work & K12 \\
WISEP J041022.71$+$150248.5  & Y0   & 1.02(0.12) & 19.61(0.04) & 19.44(0.03) & 20.02(0.05) & 19.91(0.07) &  B14   & Leg13 & C11 \\
WISE J053516.80$-$750024.9   & Y1   & $-$0.65(0.41) & 22.73(0.30) & 22.50(0.20) &   23.34(0.34)     & T14     &  M13 & this work & K12 \\
WISE J064723.23$-$623235.5   & Y1   & 0.09(0.18) & 23.13(0.09) & 22.94(0.10) & $>23.5$     & $>24$       &  K13,T14   & this work & K13\\
WISE J071322.55$-$291751.9   & Y0   & 0.18(0.08) & 20.34(0.08) & 19.98(0.05) & 20.19(0.08) & 21.30(0.31) &  T14   & this work & K12 \\
WISE J073444.02$-$715744.0   & Y0   & $-$0.66(0.19)    & 21.02(0.05) &  20.05(0.05) & 20.92(0.12) & 20.96(0.15)  &   T14     & this work &  K12 \\
WD 0806$-$66B &  Y1\tablenotemark{a}   & -1.41(0.07) & $> 23.5$ &  25.0(0.10)\tablenotemark{b} &   \nodata &   \nodata & S09 & Luh14b & Luh11 \\
WISE J085510.83$-$071442.5 & Y2\tablenotemark{a} &  3.29(0.15) &  \nodata   &  25.0($^{+0.53}_{-0.35}$)\tablenotemark{b} & \nodata &  \nodata & L\&E  & F14 & Luh14a \\
WISEPC J121756.91+162640.2B & Y0 &  $-$0.02(0.35)    & 20.26(0.03)  & 20.08(0.03) & 20.51(0.06) & 21.10(0.12) & D13 & Liu12 & Liu12\\
WISEPC J140518.40$+$553421.5 & Y0   & 0.55(0.50) & 21.24(0.10) & 21.06(0.06) & 21.41(0.08) & 21.61(0.12) & D13   & Leg13, this work & C11 \\
WISEP J154151.65$-$225025.2  & Y0.5 & 1.22(0.06) & 21.46(0.13) & 21.12(0.06) & 21.07(0.07) & 21.7(0.2)   & T14    & Leg13, this work & C11 \\
WISEP J173835.52$+$273258.9  & Y0   & 0.54(0.17) & 19.86(0.07) & 20.05(0.09) & 20.45(0.09) & 20.58(0.10) & B14    & Leg13 & C11 \\
WISEP J182831.08$+$265037.8  & Y1.5\tablenotemark{a} & 0.13(0.14) & 23.03(0.17) & 23.48(0.23) & 22.73(0.13)    & 23.48(0.36) & B14    & Leg13, this work & C11 \\
WISEPC J205628.90$+$145953.3 & Y0   & 0.73(0.13) & 19.77(0.05) & 19.43(0.04) & 19.96(0.04) & 20.01(0.06) &  B14   & Leg13 & C11 \\
WISE J222055.31$-$362817.4   & Y0   & $-$0.30(0.09) & 20.91(0.09) & 20.64(0.05) & 20.96(0.08) & 21.33(0.15) & T14    & Leg13 & K12 \\
\enddata
\tablenotetext{a}{Spectral sub-type is uncertain for WD 0806$-$66B, WISE J085510.83$-$071442.5 and
WISEP J182831.08$+$265037.8, which have very noisy or no near-infrared spectra.}
\tablenotetext{b}{The $J$ magnitude for WD 0806$-$66B has been derived from the F110W detection by Luhman et al. (2014b) as described in \S2.1. The  $J$ magnitude for WISE J085510.83$-$071442.5 is the FourStar measurement published by Faherty et al. (2014).}
\tablecomments{References are: Beichman et al. 2014; Cushing et al. 2011; Dupuy \& Kraus 2013; Faherty et al. 2014;  Kirkpatrick et al. 2011, 2012 and 2013; Leggett et al. 2013; Liu et al. 2012;  Luhman et al 2011;  Luhman 2014a; 
Luhman \& Esplin 2014, 
Luhman et al. 2014b; Marsh et al. 2013; Pinfield et al. 2014b; Subasavage et al. 2009; 
Tinney et al. 2014; 
Wright et al. 2014.}
\end{deluxetable}

\end{document}